\documentstyle[]{article}
\def\appendix#1{
\addtocounter{section}{1}
\setcounter{equation}{0}
\renewcommand{\thesection}{\Alph{section}}
\section*{Appendix \thesection\protect\indent #1}
\addcontentsline{toc}{section}{Appendix \thesection\ \ \ #1}
}

\newcommand{\0}{\frac{1}{x}}
\newcommand{\1}{\frac{1}{x-1}}

\newcommand{\5}{x}
\newcommand{\6}{(x-1)}
\newcommand{\7}{(x+\frac{n_0}{N-n_0})}
\newcommand{\8}{(x-\frac{N-n_0-n_\infty}{N-n_0})}
\newcommand{\9}{(x-\frac{n_0}{n_0-n_{\infty}})}
\newcommand{\A}{\dot{A}}

\def\a{\dot{a}}
\def\b{\dot{b}}
\newcommand{\p}{{\bf p}}
\newcommand{\k}{{\bf k}}
\newcommand{\r}{\rangle}
\def\l{\langle}

\newcommand{\q}{{\bf q}}

\newcommand{\R}{{\bf R}}
\def\D{\Delta}
\def\al{\alpha}

\def\t{\tau}
\def\T{\Theta}
\def\z{\zeta}
\def\S{\Sigma}
\def\CS{{\cal S}}
\def\e{{\,\rm e}\,}
\def\eb{{\bf e}}

\def\be{\begin{equation}}
\def\la{\label}
\def\ee{\end{equation}}
\def\bea{\begin{eqnarray}}
\def\eea{\end{eqnarray}}

\def\d{\delta}
\def\g{\gamma}
\def\G{\Gamma}
\def\U{\Upsilon}

\def\s{\sigma}
\newcommand{\cH}{{\cal H }}
\newcommand{\cL}{{\cal L }}
\newcommand{\Z}{{\bf Z}}
\begin{document}

\title{\hfill{UAHEP989} \\
\vspace{1cm}
On Lorentz invariance and supersymmetry of four particle scattering amplitudes
in $S^N\large{{\bf R}}^{8}$ orbifold sigma model}
\author{
G.Arutyunov$^{a\,c}$\thanks{arut@genesis.mi.ras.ru},
\mbox{} S.Frolov$^{b\,c}$\thanks{frolov@bama.ua.edu}
\mbox{} and \mbox{} A.Polishchuk$^{c}$\thanks{aleksey@ugcs.caltech.edu;  On
leave from Stanford University, Stanford  CA 94305-4060, USA.}
\mbox{}
\vspace{0.4cm} \\
$^a$Dipartimento di Matematica, Universita di Milano,
\vspace{-0.1cm} \mbox{} \\
"Federigo Enriques" Via C.Saldini, 50-20133 Milano, Italy
\vspace{0.4cm}
\mbox{} \\
$^b$Department of Physics and Astronomy,
\vspace{-0.1cm} \mbox{} \\
University of Alabama, Box 870324,
\vspace{-0.1cm} \mbox{} \\
Tuscaloosa, Alabama 35487-0324, USA
\vspace{0.4cm}
\mbox{} \\
$^c$Steklov Mathematical Institute,
\vspace{-0.1cm} \mbox{} \\
Gubkin str.8, GSP-1, 117966, Moscow, Russia
\mbox{}
}
\date {}
\maketitle
\begin{abstract}
The $S^N\R^{8}$ supersymmetric orbifold sigma model is expected to describe
the IR limit of the Matrix string theory. In the framework of the model the
type IIA string interaction is governed by a vertex which was recently 
proposed by R.Dijkgraaf, E.Verlinde and H.Verlinde. By using this 
interaction vertex we derive all four particle scattering amplitudes directly 
from the orbifold model in the large $N$ limit.
\end{abstract}

\section{Introduction}
To provide a heuristic basis for understanding various phenomena 
arising in superstrings, it was suggested that there exists a fundamental 
nonperturbative quantum theory in eleven dimensions, called M-theory. 
The appropriate compactification of M-theory leads to one of the five 
supersting theories and, in particular, the compactification on 
$S^1$ leads to the ten-dimensional type IIA superstring theory \cite{W2,Sch}. 
Although at present we do not know how to formulate M-theory as a quantum 
theory it has been conjectured \cite{BFSS} that there is a precise 
equivalence between the M-theory and the large $N$ limit of the 
supersymmetric quantum matrix model which describes the dynamics of 
D-particles \cite{W1}.

In the original $D$-particle language, $S^1$ compactification of $M$-theory
amounts to applying a $T$-duality transformation along the $S^1$ direction,
thereby turning the D-particles into D-strings. By adopting this approach
we can cast matrix theory into the form of the two-dimensional
${\cal{N}} = 8$ maximally supersymmetric $U(N)$ Yang-Mills theory \cite{T}.
According to the matrix theory philosophy, in the limit $N \to \infty$ the
Yang-Mills theory should describe nonperturbative dynamics of type IIA
superstrings. This is a new type of nonperturbative duality between a gauge 
theory and a string theory in which the string coupling constant is 
inversely proportional to the YM coupling constant: 
$g_{YM}^{-2} = \alpha' g_s^{2}$ \cite{M,BS,DVV}. Thus, we expect that 
the strong coupling expansion of the Yang-Mills model describes
the perturbative type IIA free string theory $(g_s=0)$.
Recently, it was conjectured by R.Dijkgraaf, E.Verlinde and 
H.Verlinde \cite{DVV} that in the IR limit the Yang-Mills model reduces to 
the ${\cal{N}} = 8$ nonabelian $S^N \R^{8}$ supersymmetric orbifold
sigma model. The fact that the orbifold model is nonabelian comes as no
surprise since in the IR limit the original gauge symmetry group $U(N)$
reduces to the permutation group $S_N$. Furthermore, in \cite{DVV} it was
proposed that the string interaction in the orbifold sigma model is governed 
by a supersymmetric vertex of conformal dimension $(\frac32,\frac32)$. 
This vertex describes the elementary process of joining and splitting of 
strings and from the viewpoint of the gauge theory is responsible for 
partial restoring of the $U(N)$ gauge symmetry in some small region of 
space-time.
With the DVV interaction vertex at hand one is tempted to deduce string
scattering amplitudes directly from the orbifold sigma model.
It should be realized that this is a nontrivial problem due to the nonabelian 
nature of the orbifold.
Nevertheless, the necessary tools for computing tree-level diagrams were
recently developed in \cite{AF,AF2}. In particular,
the four-graviton scattering amplitudes for type IIA and IIB strings were
calculated and were shown to be Lorentz invariant in the large $N$ limit.
It was also observed that the string kinematical factor exhibited manifest 
Lorentz invariance even at finite $N$.

In this paper we complete the proof of the DVV conjecture on the
level of tree diagrams by explicitly calculating all four particle
scattering amplitudes for type IIA superstrings directly from the $S^N \R^8$
supersymmetric orbifold sigma model and demonstrating their Lorentz and 
supersymmetry invariance. This provides a new consistency check on the 
matrix model conjecture. Furthermore, this is a new evidence of
the hidden supersymmetry invariance of the matrix model and its existence
is a necessary condition for the model to describe M-theory.

We begin by reviewing the general formalism of the $S^N \R^8$ supersymmetric
orbifold sigma model, developed in \cite{AF2}. We define the $S_N$
invariant vertex operators which create all massless states of type IIA
string theory and which form a closed operator algebra. Following the approach
of
\cite{DVV}, we describe the DVV interaction vertex which is both space-time
supersymmetric and $SO(8)$ invariant.
Then, we construct the $S$-matrix to the second order in the
coupling constant by sandwiching two DVV
vertices in-between the asymptotic states corresponding to two incoming
and two outgoing particles. As a result of this construction 
we obtain the expression for the $S$-matrix element as the sum over specific 
four-point correlation functions which we explicitly list at the end of 
Section 2.3. The procedure for calculating these correlation functions 
was outlined in \cite{AF2} and in Section 2.4 we summarize the main results.
The appropriate scattering amplitude can then be obtained from the 
$S$-matrix element by making use of the reduction formula. Since the problem
of calculating scattering amplitudes is equivalent to that of calculating 
all possible open string kinematical factors it follows that to prove 
the DVV conjecture on the level of tree diagrams we have to show that all 
kinematical factors obtained directly from the orbifold sigma model coincide 
with those obtained in the framework of the superstring theory. To this end, 
we first compute the open string kinematical factor
corresponding to the scattering of two vector particles and two fermions. In 
the process of this calculation we develop the necessary tools to deal with 
spinors and focus on the issue of the Lorentz invariance of the model. It 
turns out that the kinematical factor that we obtain is automatically Lorentz 
invariant and coincides with the well-known open string kinematical factor
of the superstring theory. We then compute the remaining kinematical factors 
for all massless particles which make up the complete spectrum of IIA 
supergravity and show that they also coincide with those of the superstring 
theory. In conclusion we discuss interesting problems that still remain open.

\section{General Formalism}
\subsection{Free $S^N \R^8$ orbifold model}
\setcounter{equation}{0}
The action that defines
the free $S^N\R^{8} \equiv (\R^8)^N/S_N$ orbifold sigma model is
\be S=\frac{1}{2\pi}\int d\tau d\sigma
\sum_{I=1}^{N}(\partial_{\tau}X^i_I\partial_{\tau}X^i_I
-\partial_{\sigma}X^i_I\partial_{\sigma}X^i_I +
\frac{i}{2}\theta^{a}_I(\partial_{\tau}+\partial_{\sigma})\theta^{a}_I+
\frac{i}{2}\theta^{\a}_I(\partial_{\tau}-\partial_{\sigma})\theta^{\a}_I).
\label{act}
\ee
Here $X^i$ are eight real bosonic fields transforming in the
${\bf 8_v}$ representation of the transversal group $SO(8)$ and
$\theta^{a},\theta^{\a}$ $a,\a=1,\ldots ,8$ are sixteen fermionic fields
transforming in the ${\bf 8_s}$ and ${\bf 8_c}$ representations respectively.
As pertains to all orbifold models \cite{DHVW1,DHVW2} the fundamental fields
$X^i$ and $\theta^{\alpha}$ are allowed to have twisted boundary conditions:
\be
X^i(\s +2\pi )=gX^i(\s ), \qquad
\theta^{\alpha}(\s +2\pi )=g\theta^{\alpha}(\s ),
\label{twistbc}
\ee
where in the case of the $S^N \R^{8}$ orbifold model $g \in S_N$.

In the conventional QFT the scattering amplitude
to the second order in the coupling constant
is extracted from the $S$-matrix element, schematically written as
$$
\l f| S | i \r \sim \l f | \int d x_1 d x_2 T \{ V_{int}(x_1) V_{int}(x_2) \} 
| i \r
$$
by using the reduction formula.
Consequently, to compute scattering amplitudes we first need to define
{\it in} $(| i \r)$ and {\it out} $(| f \r)$ states which are the states 
in the Hilbert space of the
$S^N\R^8$ orbifold sigma model. Recall that the Hilbert space of an orbifold
model decomposes into the
direct sum of Hilbert spaces of twisted sectors 
corresponding to conjugacy classes of the
discrete group defining the orbifold. The
conjugacy classes of $S_N$ are described by partitions
$\{ N_n\}$ of $N$ and can be represented by
\be
[g] = (1)^{N_1}(2)^{N_2} \cdots (s)^{N_s}, \quad  N= \sum_{n=1}^s n N_n,
\label{facg}
\ee
\noindent where $N_n$ is the multiplicity of the cyclic permutation
$(n)$ of $n$ elements.
In any conjugacy class $[g]$ there is only one element $g_c$ that has
the canonical block-diagonal form
\bea
\nonumber
&&
g_c = diag(\underbrace{\omega_1,\ldots,\omega_1}_{N_1\ times},
           \underbrace{\omega_2,\ldots,\omega_2}_{N_2\ times}, \ldots
           \underbrace{\omega_s,\ldots,\omega_s}_{N_s\ times}),
\eea
where $\omega_n$ is an $n \times n$ matrix that generates the cyclic
permutation $(n)$ of $n$ elements. Since $\omega_n$ generates the group $\Z_n$,
as can be easily verified, the
Hilbert space $\cH_{[g]}\equiv\cH_{\{ N_n\}}$
is decomposed into the graded $N_n$-fold symmetric tensor products
of Hilbert spaces $\cH_{(n)}$ which are $\Z_n$ invariant subspaces of the
Hilbert space:
\bea
{\cH}_{\{N_n\}} = \bigotimes_{n=1}^s \, S^{N_n} \cH_{(n)}
=\bigotimes_{n=1}^s \,\left(\underbrace{\cH_{(n)} \otimes \cdots
\otimes \cH_{(n)}}_{N_n\ times}\right)^{S_{N_n}}.
\nonumber
\eea
The fundamental fields corresponding to the space ${\cH}_{(n)}$ are
$8n$ bosonic fields $X_I^i$ and $16n$ fermionic fields
$\theta^{\alpha}$ with the cyclic boundary condition
\be
\la{cyc}
X_I^i(\s +2\pi )=X_{I+1}^i(\s), \qquad
\theta_I^{\alpha}(\s +2\pi )=\theta_{I+1}^{\alpha}(\s), \qquad I=1,2,\ldots,n.
\ee
As usual, states of the Hilbert space $\cH_{(n)}$ are obtained by acting
on momentum eigenstates with the string creation operators. Since the
fundamental
fields have twisted boundary conditions, the string creation operators have
nontrivial transformation properties under the action of the group $S_N$.
However, the space $\cH_{(n)}$ must be $\Z_n$ invariant and to ensure
this one has to impose  the condition on the allowed states of $\cH_{(n)}$:
\bea
(L_0-\bar {L}_0)|\Psi\rangle =nm|\Psi\rangle ,
\nonumber\eea
where $m$ is some integer and $L_0$ is the canonically normalized
$L_0$-operator of a single long string obtained by glueing together the fields
$X_I(\s)$ $(\theta_I(\s))$ into one field $X(\s)$ $(\theta(\s))$.

Before passing on to the construction of asymptotic states corresponding to
${\cH}_{(n)}$, we note that according to \cite{DMVV}
the Fock space of the second-quantized IIA type
string is recovered in the limit $N\to\infty$,
$\frac {n_i}{N}\to p^+_i$, where the finite ratio $\frac {n_i}{N}$
is identified with the $p^+_i$ momentum of a long string. In this limit the
$\Z_n$ projection becomes the usual level-matching condition
$L_0^{(i)}-\bar {L}_0^{(i)}=0$ for closed strings, while the individual
$p^-_i$ light-cone momentum is defined by means of the standard mass-shell
condition $p^+_ip^-_i=L_0^{(i)}$.

\subsection{Asymptotic states of $S^N\R^{8}$}
We will consider the conformal field theory on the sphere with coordinates
$(z,\bar z)$ obtained from the cylinder with coordinates $(\t,\s)$
by performing the Wick rotation $\tau\to -i\tau$ followed by the
map: $z=\e ^{\tau +i\s }$, $\bar {z}=\e ^{\tau -i\s }$.

The asymptotic states of the orbifold CFT model are
obtained by acting with the $S_N$-invariant vertex operators on the NS vacuum
$|0\r$ which is normalized according to
$$
\l 0|0 \r = R^{8N}.
$$
Here $R$ is the radius of a circle onto which we compactify the string
coordinates $x^i_I$ in order to regularize the sigma model.

The most natural way to build $S_N$-invariant vertex
operators $V_{[g]}$ is to first introduce a vertex operator $V_{g}$
corresponding to a particular group element $g$ of $S_N$ and
then sum over the conjugacy class of $g$. This procedure can
be represented as follows:

\be
\la{opdef}
V_{[g]}(z,\bar z) = \frac{1}{N!}\sum_{h \in S_N} V_{h^{-1}gh}(z,\bar z).
\ee

The vertex operators $V_g(z,\bar z)$ should be constructed
from the twist fields of the orbifold model - the fields about which 
the fundamental fields have twisted boundary conditions.
Since the monodromy conditions of the bosonic fundamental fields
$X^i_I(z,\bar z)$ are given by
eq.(\ref{cyc}), we are led to the following definition of the bosonic
twist field $\s_g(z,\bar z)$:
$$
X^i(z e^{2 \pi i}, \bar z e^{-2 \pi i}) \s_g(0,0) =
g X^i(z, \bar z) \s_g(0,0)
$$
In exactly the same manner we introduce the fermionic twist field
$\S_g(z,\bar z)$.

In constructing the vertex operator $V_g(z,\bar z)$ one is tempted to
consider the tensor product of the bosonic twist field $\s_g(z,\bar z)$ and 
the fermionic twist field $\S_g(z,\bar z)$.
Although the nonabelian nature of the orbifold sigma model does not admit the
factorization into bosonic and fermionic (holomorphic and antiholomorphic)
contributions, it was shown in \cite{AF2} that
this factorization can be assumed provided that one introduces a certain
normalization constant, later denoted by $\kappa$, at the final stage of
scattering amplitude calculation.
Thus, we define the vertex operator $V_g(z, \bar z)$ according to
\be
\la{vertex}
V_g(z,\bar z) = \s_g(z) \S_g(z) {\bar \s}_g(\bar z) {\bar \S}_g(\bar z).
\ee

To clarify the meaning of the holomorphic (anti-holomorphic) twist field
$\s_g(z)$ ($\bar{\s}_g(\bar z)$)
we decompose the fundamental field $X(z,\bar z)$ into the left- and
right-moving components:
$$
2X(z,\bar z) = X(z) + X(\bar z),
$$
so that now we can define $\s_g(z)$ and
$\bar{\s}_g(\bar z)$ according to
$$
X^i(z e^{2 \pi i}) \s_g(0) =
g X^i(z) \s_g(0) 
\quad \Leftrightarrow \quad
X^i(z e^{2 \pi i}) \s_{g^{-1}}(0) =
g^{-1} X^i(z) \s_{g^{-1}}(0)
$$
and
$$
\bar{X}^i(\bar z e^{- 2 \pi i}) \bar{\s}_g(0) =
g \bar{X}^i(\bar z) \bar{\s}_g(0) 
\quad \Rightarrow \quad
\bar{X}^i (\bar z e^{2 \pi i}) \bar{\s}_{g}(0) =
g^{-1} \bar{X}^i (\bar z) \bar{\s}_{g}(0).
$$

\noindent Now the formal substitution $z \to \bar{z}$ leads to the
conclusion that the
operator $\s_g$ is identical to the operator $\bar{\s}_{g^{-1}}$.

\noindent For any element $g \in S_N$ with the decomposition
\be
\la{gr.el}
g = (n_1)(n_2) \cdots (n_{N_{str}})
\ee
we represent $V_g(z,\bar z)$ as the tensor product of operators each
corresponding to some cycle $(n_{\alpha})$:
$$
V_g(z,\bar z) = \bigotimes_{\al =1}^{N_{str}} \, V_{(n_\al)}(z,\bar z).
$$
The operator, $\s_{(n)}(z,\bar z) = \s_{(n)}(z) \bar{\s}_{(n)}(\bar z)$ is a
primary field
\cite{BPZ} that creates the bosonic vacuum state of a twisted sector, labeled
by $(n)$, at the point $(z,\bar z)$. We denote
this vacuum state by $|(n) \r = \s_{(n)}(0,0)|0\r$. Recall that zero modes of
fundamental fields $\theta^{\alpha}$ form the Clifford algebra. Therefore,
by triality the vacuum state can be chosen to be the direct sum 
${\bf 8_v} \bigoplus {\bf 8_c}$.
Consequently, we define the primary spin fields of the holomorphic sector
$\S_{(n)}^i(z)$ , $\S_{(n)}^{\a}(z)$ which create the fermionic vacuum state:
$|(n),\dot \mu \r = \S_{(n)}^{\dot \mu}(0)|0\r$, where $\dot \mu = (i,\a)$.
Under the world-sheet parity
$z \to \bar{z}$
and the space reflection $X^3\to -X^3$ twist fields transform as follows
\bea
\nonumber
&& \s_{(n)}(z) \leftrightarrow {\bar \s}_{(-n)}(\bar{z}); \qquad
\Sigma^{\a}_{(n)}(z)\leftrightarrow\bar{\Sigma}^{a}_{(-n)}(\bar{z});\\
\label{trans}
&&\Sigma^i_{(n)}(z)\leftrightarrow\bar{\Sigma}^i_{(-n)}(\bar{z}), \qquad
i\neq 3; \qquad  \Sigma^3_{(n)}(z)\leftrightarrow
-\bar{\Sigma}^3_{(-n)}(\bar{z}),
\eea
where $(-n)$ denotes the cycle with the reversed orientation corresponding to
the element $\omega_n^{-1}$. The third direction is singled out since in our
conventions $\g^3=1$ (see Appendix A).

Finally, we introduce the primary field $\s_g[\{ \k_\alpha \}](z,\bar z)$
corresponding to particles with transversal momenta $\k_\alpha$.
Suppose that $g \in S_N$ has the decomposition (\ref{gr.el}) so that
the following factorization takes place
$$
\s_g(z,\bar z) = \bigotimes_{\alpha=1}^{N_{str}} \s_{(n_\alpha)}(z,\bar z),
$$
then $\s_g[\{\k_\alpha \}](z,\bar z)$ is defined by
$$
\s_g[\{\k_\alpha\}](z,\bar z) =
: e^{i \frac{k^i_{\alpha}Y^i_{\alpha}(z,\bar z)}{\sqrt{n_\alpha}}} :
\s_g(z,\bar z) \equiv
\bigotimes_{\alpha = 1}^{N_{str}} \s_{(n_\alpha)}[\k_\alpha],
$$
where $n_1=n_2=\cdots=n_{N_1} = 1, n_{N_1+1}=n_{N_1+2}=\cdots=n_{N_1+N_2}=2,
\ldots$ and
$$
Y^i_{\alpha}(z,\bar z) = \frac{1}{\sqrt{n_\alpha}} \sum_{I=1}^{n_\alpha}
X^i_I(z,\bar z).
$$
Combining the fermionic vacuum state with the vacuum state of the bosonic 
sector we find 256 states that describe the complete spectrum of type 
IIA supergravity. In particular, the state with $k^+ = \frac{n}{N}$, 
transversal momentum $\k$ and polarization $\z^{\dot \mu \nu}$\footnote{In 
what follows we call the wave function of a particle a polarization.} 
is generated from the NS vacuum $|0\r$ by the vertex operator
\be
\la{invver}
V_{(n)}[\k,\z](z,\bar z) = \z^{\dot \mu \nu} \s_{(n)}[\k](z,\bar{z})
\Sigma^{\dot{\mu}}_{(n)}(z) \bar{\Sigma}^{\nu}_{(n)}(\bar{z}).
\ee

\noindent As was shown in \cite{AF2} $S_N$-invariant vertex operators
\be
\la{invver_1}
V_{[g]}[\{\k_\alpha, \z_\alpha \}] =  \frac{1}{N!}
\sum_{h \in S_N} \bigotimes_{\alpha=1}^{N_{str}}
V_{h^{-1}(n_\alpha)h}[\k_\alpha,\z_\alpha]
\ee
creating ground states, i.e. states with $\k_{\al}\equiv 0$, have the same 
conformal dimension which is a necessary condition for the orbifold
sigma model to originate from the IR limit of the Yang-Mills theory.

Next we turn to the description of the DVV interaction vertex. To this end
we introduce the first excited state $\t_{(n)}(z,\bar z)$ of the twisted sector 
which appears as the most singular term in the OPE
\be
\la{tau}
\partial X_I^i(z) \s_{(n)}(w)=(z-w)^{-\left(1-\frac{1}{n}\right)}
e^{\frac{2\pi i}{n}I}\tau_{(n)}^i(w)+\ldots \quad .
\ee
Suppose  $(n)$ is a simple transposition ($n=2$) which exchanges
$X_I$ with $X_J$, then we can define the field $\t_{IJ} \equiv \t_{(2)}$.
The DVV interaction vertex \cite{DVV} is then given by
\bea
V_{int}=-\frac{\lambda N}{2\pi } \sum_{I<J}\int d^2z|z| \left(\t^i(z)\S^i(z)
\bar{\t}^j(\bar z)\bar{\S}^j(\bar z)\right)_{IJ},
\la{DVV}
\eea
where $\lambda$ is a coupling constant proportional to the string coupling
$g_s$.

The twist field $V_{IJ}(z,\bar z) \equiv \left(\t^i(z)\S^i(z)
\bar{\t}^j(\bar z)\bar{\S}^j(\bar z)\right)_{IJ}$
is a weight
$(\frac32 ,\frac32)$ conformal field and the coupling constant
$\lambda$ has dimension $-1$. As was shown in \cite{DVV} this
interaction vertex is space-time supersymmetric, $SO(8)$ invariant
and describes the elementary string interaction. In addition, it is invariant
with respect to the world-sheet parity transformation $z \to \bar z$ and
an odd number of space reflections.

\subsection{S-matrix element}
With the account of (\ref{DVV}) the S-matrix element to the second order in 
the coupling constant $\lambda$ is given by the formula
\be
\langle f|S|i\rangle = -\frac12 \left(\frac{\lambda N}{2\pi }\right)^2
\langle f|\int d^2z_1d^2z_2|z_1||z_2|T\left(
\cL_{int}(z_1,\bar z_1)
\cL_{int}(z_2,\bar z_2)\right) |i\rangle ,
\label{matel}
\ee
where $T$ means time-ordering: $|z_1|>|z_2|$ and
\bea
\cL_{int}(z,\bar z)=\sum_{I<J} V_{IJ}(z,\bar z).
\nonumber\eea

For the initial state $|i\rangle$ we choose the state corresponding to
two incoming particles with transversal momenta $\k_1$ and $\k_2$ and 
polarizations $\z_1$ and $\z_2$ and for the final state $\l f|$ - the sate
corresponding to two outgoing particles with transversal momenta $\k_3$ and 
$\k_4$ and polarizations $\z_3$ and $\z_4$, respectively:
\bea
&&
|i\rangle = C_0 V_{[g_0]}[\k_1,\z_1;\k_2,\z_2](0,0)|0\rangle , \\
\nonumber
&&
\langle f|= C_\infty \lim_{z_\infty\to\infty}|z_\infty |^{4\D_\infty}
\langle 0|V_{[g_\infty]}[\k_3,\z_3;\k_4,\z_4](z_\infty ,\bar z_\infty ).
\eea
\noindent Recall that $S_N$ invariant vertex operators
$V_{[g]}[ \{ \k_\alpha,\z_\alpha \} ](z,\bar z)$
were defined in (\ref{invver_1}). The elements $g_0$, $g_\infty$ are chosen in
the canonical block-diagonal form
$$
g_0=(n_0)(N-n_0), \qquad g_\infty =(n_\infty)(N-n_\infty)
$$
and to ensure proper normalization the constants $C_0$ and $C_\infty$ have to 
be equal to
$$
C_0=\sqrt {\frac {N!}{n_0(N-n_0)}},\quad
C_\infty =\sqrt {\frac {N!}{n_\infty (N-n_\infty )}}.
$$
Following the approach of \cite{DVV} we introduce the light-cone
momenta of initial and final particles
\bea
&&
k_1^+=\frac{n_0}{N},\quad
k_2^+=\frac{N-n_0}{N},\quad
k_3^+=-\frac {n_\infty}{N},\quad
k_4^+=-\frac {N-n_\infty}{N},
\nonumber
\eea
which satisfy the mass-shell condition: $k_a^+k_a^- - \k_a\k_a = 0$ for
each $a$, where $a=1,\ldots, 4$.
\noindent According to \cite{AF2}
the $S$-matrix element can be written as
\be
\la{S}
\l f | S |i \r = -i2\lambda^2N^3 \d(k_1^-+k_2^-+k_3^-+k_4^-)\, {\cal M},
\ee
where the delta function results from the
integral over $z_1$  and ${\cal M}$ is given by
\be
\la{M}
{\cal M} = \int d^2u |u| F(u,\bar u).
\ee
\noindent Here we introduced a concise notation
\bea
\la{fuu}
&&
F(u,\bar u) = \langle f|T\left( \cL_{int}(1,1)
\cL_{int}(u,\bar u)\right) |i\rangle  \\
\nonumber
&&
= C_0 C_\infty  \sum_{I<J;K<L}
\l V_{[g_{\infty}]}[\k_3,\z_3;\k_4,\z_4](\infty) T \left( V_{IJ}(1,1)
V_{KL}(u,\bar u) \right) V_{[g_0]}[\k_1,\z_1;\k_2,\z_2](0,0)\r.
\eea
In what follows we assume for definiteness that $|u| < 1$.
From the definition (\ref{opdef}) of $V_{[g]}$ it is clear that (\ref{fuu})
is the sum over two conjugacy classes corresponding to
group elements $g_0$ and $g_\infty$. However, with the account of the
invariance of
the interaction vertex as well as of any correlation function constructed
from vertex operators under the global action of the symmetric group it 
becomes possible to reduce the sum over two conjugacy classes to the 
single sum:
\be
\la{fuu_1}
F(u,\bar u) =\frac{C_0 C_\infty}{N!} \sum_{h_\infty \in S_N} \sum_{I<J;K<L}
\l V_{h_{\infty}^{-1}g_{\infty}h_{\infty}}(\infty)V_{IJ}(1,1)V_{KL}(u,\bar u)
V_{g_0}(0,0)\r.
\ee
The obtained expression can be further simplified, however, to do so we need
to establish certain properties of correlation functions entering (\ref{fuu_1}).
To this end we recall that the action (\ref{act}) and the
DVV interaction vertex are invariant under the world-sheet parity
transformation $z\to \bar z$ combined with the space reflection $X^3~\to~-X^3$,
while the vertex operator
$V_g[\{ \k_{\alpha},\z_{\alpha} \}](z,\bar z)$ transforms into
$ {\tilde V}_{g^{-1}}[\{ \k_{\alpha},\z_{\alpha} \}](z,\bar z) \equiv
V_{g^{-1}}[\{ \tilde{\k}_{\alpha},\tilde{\z}_{\alpha} \}](z,\bar z)$,
where $\tilde{k}_\alpha$ , $\tilde{\z}_\alpha$ are the space reflected
momenta and polarization respectively, $\tilde{k}^3=-k^3$.
Let us consider the correlation function
$\l V_{h_{\infty}^{-1} g_{\infty} h_{\infty}} V_{IJ} V_{KL} V_{g_0} \r$
with the monodromy condition
$$
h_{\infty}^{-1} g_{\infty} h_{\infty} g_{IJ} g_{KL} g_0 =1 \quad
\Rightarrow \quad h_{\infty}^{-1} g_{\infty} h_{\infty} = g_0^{-1} g_{KL} g_{IJ} .
$$
With the account of the world-sheet parity and the space reflection
symmetries we obtain the following equality:
$$
\l V_{g_0^{-1} g_{KL} g_{IJ}} V_{IJ} V_{KL} V_{g_0} \r =
\l {\tilde V}_{g_{IJ} g_{KL} g_0} V_{IJ} V_{KL} {\tilde V}_{g_0^{-1}}\r
$$
Due to the invariance of the correlation function under
the global action of $S_N$ and the fact that the elements $g$ and $g^{-1}$
belong to the same conjugacy class we obtain
$$
\l V_{g_0^{-1} g_{KL} g_{IJ}} V_{IJ} V_{KL} V_{g_0} \r =
\l {\tilde V}_{g_{I'J'} g_{K'L'} g_0^{-1}} V_{I'J'} V_{K'L'} {\tilde V}_{g_0}\r
$$
where $g_{I'J'} = h g_{IJ} h^{-1} \,\, , \,\, g_{K'L'}=hg_{KL} h^{-1}$, and
the element
$h$ is such that $g_0^{-1} = h^{-1} g_0 h$. Due to the $SO(8)$ invariance of
the model the correlation function (\ref{fuu}) 
can depend only on the scalar products of momenta $\k_{\alpha}$ and 
polarizations $\z_{\alpha}$ as well as on their contractions with the $SO(8)$ 
spin-tensor $\g^{ij}_{\a \b}$. Obviously all scalar products are invariant 
under the space reflection while $\g^{ij}_{\a \b}$ transforms into
${\tilde \g}^{ij}_{ab}$. Here ${\tilde \g}^{i} = \g^i$ for
$i \neq 3$ and ${\tilde \g}^{3} = - \g^3$. From the explicit form of
$\g^{ij}_{\a \b}$ given in Appendix A and with the account of
$\g^{ij}_{a b} \equiv ({\g^{ij}}^{T})_{\a \b}$, one can easily deduce that
$$
\g^{ij}_{\a \b} =
{\tilde \g}^{ij}_{ab}.
$$
Thus we are justified to make the replacement 
$\tilde{\k}_{\alpha} \to \k_{\alpha}$ and $\tilde{\z}_{\alpha} \to \z_{\alpha}$
in the correlation function. Consequently, we arrive at the equality
\be
\la{doubling}
\l V_{g_0^{-1} g_{KL} g_{IJ}} V_{IJ} V_{KL} V_{g_0} \r =
\l V_{g_{I'J'} g_{K'L'} g_0^{-1}} V_{I'J'} V_{K'L'} V_{g_0}\r .
\ee
Now note that while the correlation function on the left hand side of (\ref{doubling}) corresponds to
$\l V_{h_{\infty}^{-1} g_{\infty} h_{\infty}} V_{IJ} V_{KL} V_{g_0} \r$
with the monodromy condition
$$
h_{\infty}^{-1} g_{\infty} h_{\infty} g_{IJ} g_{KL} g_0 = 1
$$
the correlation function on the right hand side of eq.(\ref{doubling})
satisfies the monodromy condition
$$
h_{\infty}^{-1} g_{\infty} h_{\infty} = g_{I'J'} g_{K'L'} g_0^{-1} \quad
\Rightarrow \quad
h_{\infty}^{-1} g_{\infty} h_{\infty} g_0 g_{K'L'} g_{I'J'} = 1.
$$
Therefore, the contribution of terms satisfying either of the two monodromy 
conditions coincide. As it was shown in \cite{AF2} the only
nontrivial terms in (\ref{fuu_1}) are those that satisfy precisely
these two monodromy conditions. Consequently, we can include only terms
corresponding to one of the monodromy conditions and place a factor of 2 in
front of the entire expression. Using the same procedures as those in
establishing (\ref{doubling}) we now show that the correlation function 
$F(u,\bar u)$ is real. To this end we first consider the result of complex 
conjugating the correlation function:
\bea
\nonumber
&&
{\l V_{g_\infty}[\k_3,\z_3;\k_4,\z_4](\infty)V_{IJ}(1,1)V_{KL}(u,\bar u)
V_{g_0}[\k_1,\z_1;\k_2,\z_2](0,0)\r}^* \\
\nonumber
&&=
\lim_{ z_{\infty} \to \infty } \lim_{z_0 \to 0}
{|z_{\infty}|}^{ -4 \D_{g_\infty} [ \{ \k_3, \k_4 \} ]}
{|z_{0}|}^{-4 \D_{g_0} [ \{ \k_1, \k_2 \} ]}
\, |u|^{-6} \\
\nonumber
&& \times
\l V_{g_0^{-1}}[-\k_1,\z_1;-\k_2,\z_2]
(\frac{1}{z_\infty},\frac{1}{\bar{z}_\infty})
V_{KL}(\frac{1}{u},\frac{1}{\bar u})V_{IJ}(1,1)
V_{g_{\infty}^{-1}}[-\k_3,\z_3;-\k_4,\z_4](\frac{1}{z_0},\frac{1}{\bar{z}_0})\r ,
\eea
where we took into account the conjugating property of a vertex operator
\be
\la{dagger}
(V_g[\{\k_{\alpha}\}](z))^{\dagger} =
z^{-2 \D_g[\{ \k_\alpha \}]} V_{g^{-1}}[\{-\k_{\alpha}\}](\frac{1}{z}),
\ee
and the fact that the DVV vertex is of conformal dimension
$(\frac32,\frac32)$. Due to the $SO(8)$ invariance we can make
a replacement $-\k_{\alpha} \to \k_{\alpha}$ and after performing the conformal
transformation $z \to \frac{1}{z}$ obtain:
\bea
\nonumber
&&
{\l V_{g_\infty}[\k_3,\z_3;\k_4,\z_4](\infty)V_{IJ}(1,1)V_{KL}(u,\bar u)
V_{g_0}[\k_1,\z_1;\k_2,\z_2](0,0)\r}^* \\
\nonumber
&&=
\l V_{g_\infty^{-1}}[\k_3,\z_3;\k_4,\z_4](\infty)
V_{IJ}(1,1)V_{KL}(u,\bar u)V_{g_0^{-1}}[\k_1,\z_1;\k_2,\z_2](0,0)\r \\
\nonumber
&&=\l V_{{g'}_\infty}[\k_3,\z_3;\k_4,\z_4](\infty)
V_{I'J'}(1,1)V_{K'L'}(u,\bar u)V_{g_0}[\k_1,\z_1;\k_2,\z_2](0,0)\r  ,
\eea
where $h \in S_N$ is the solution of $h^{-1} g_0^{-1} h = g_0\,$ and
$$
h^{-1} g^{-1}_{\infty} h = g^{'}_{\infty}, \quad h^{-1} g_{IJ} h = g_{I'J'},
\quad h^{-1} g_{KL} h = g_{K'L'}.
$$
Now we apply this result to find the complex conjugate of $F(u,\bar u)$:
\bea
\nonumber
&&
{F(u,\bar u)}^* =
\frac{2 C_0 C_\infty}{N!} {\sum_{h_\infty \in S_N}}^{'} \sum_{I<J;K<L}
{\l V_{h_{\infty}^{-1}g_{\infty}h_{\infty}}[\k_3,\z_3;\k_4\z_4](\infty)
V_{IJ}(1,1)V_{KL}(u,\bar u)V_{g_0}[\k_1,\z_1;\k_2,\z_2](0,0)\r}^* \\
\nonumber
&&=
\frac{2 C_0 C_\infty}{N!} {\sum_{h_\infty \in S_N}}^{'} \sum_{I<J;K<L}
\l V_{{h'}_{\infty}^{-1} g_{\infty} {h'}_{\infty}}[\k_3,\z_3;\k_4\z_4]
(\infty) V_{I'J'}(1,1)V_{K'L'}(u,\bar u)V_{g_0}[\k_1,\z_1;\k_2,\z_2](0,0)\r \\
\nonumber
&&=
\frac{2 C_0 C_\infty}{N!} {\sum_{{h'}_\infty \in S_N}}^{'} \sum_{I'<J';K'<L'}
\l V_{{h'}_{\infty}^{-1} g_{\infty} {h'}_{\infty}}[\k_3,\z_3;\k_4\z_4]
(\infty) V_{I'J'}(1,1)V_{K'L'}(u,\bar u)V_{g_0}[\k_1,\z_1;\k_2,\z_2](0,0)\r \\
\nonumber
&&
=F(u,\bar u),
\eea
where
$$
h^{-1} g_0^{-1} h = g_0 , \quad
h^{-1} h^{-1}_{\infty} g_{\infty}^{-1} h_{\infty}h = {h'}^{-1}_{\infty}
g_{\infty} {h'}_{\infty}
$$
and the prime in the sum over $h_{\infty}$ indicates that we include
only terms which satisfy the monodromy condition
$h^{-1}_{\infty} g_{\infty} h_{\infty} g_{IJ} g_{KL} g_0 = 1$.
This completes the proof.

As was shown in \cite{AF2}, using the global $S_N$ invariance
of the model one can recast $F(u,\bar u)$ into the following form
\bea
\nonumber
&&
F(u,\bar u) = 2 N^2 \sqrt{k_1^+k_2^+k_3^+k_4^+}
\left(\sum_{I=1}^{n_\infty}\l V_{g_\infty (I)}
(\infty) V_{I,I+N-n_\infty}(1,1)V_{n_0N}(u,\bar u)V_{g_0}
(0,0)\r
\right. \\
\nonumber
&&
+\sum_{I=1}^{N-n_\infty}\langle V_{g_\infty (I)}
(\infty) V_{I,I+n_\infty}(1,1)V_{n_0 N}(u,\bar u)V_{g_0}
(0,0)\r +
\sum_{J=n_0+1}^{n_\infty}
\langle V_{g_\infty (J)}
(\infty) V_{n_0J}(1,1)V_{n_\infty N}(u,\bar u)V_{g_0}
(0,0)\r
\nonumber\\
&&+\left.\sum_{J=n_0+n_\infty +1}^{N}
\langle V_{g_\infty (J)}
(\infty) V_{n_0J}(1,1)V_{n_0+n_\infty ,N}(u,\bar u)V_{g_0}
(0,0)\r
\right),
\la{fuu1}
\eea
where the elements $g_\infty $ have to be found from the equation
$g_\infty g_{IJ}g_{KL}g_0=1$\footnote{Here we assume for definiteness that
$n_0<n_\infty$.}. To simplify the notation we did not
explicitly indicate the momenta $\k$ and polarizations $\z$ in (\ref{fuu1}).

Consequently, the $S$-matrix element is constructed from the correlation
functions
\be
\la{guu}
G_{IJKL}(u,\bar u) \equiv
\l V_{g_\infty}[\k_3,\z_3;\k_4,\z_4](\infty) V_{IJ}(1,1) V_{KL}(u,\bar u)
V_{g_0}[\k_1,\z_1;\k_2,\z_2](0,0)\r
\ee
corresponding to $|u| < 1$ and the correlation functions obtained from
(\ref{guu}) by interchanging $(u,\bar u) \leftrightarrow (1,1)$ and therefore
corresponding to $|u| > 1$.
Here all possible combinations of $g_{\infty}$ , $g_{IJ}$ and $g_{KL}$,$g_0$
are listed in (\ref{fuu1}).

\subsection{Correlation functions}
Taking into account the definition (\ref{invver}) of
$V_g[\k_\alpha,\z_\alpha]$
and the expression (\ref{DVV}) for the DVV interaction vertex
we obtain the holomorphic contribution to the correlation function (\ref{guu}):
$$
G_{IJKL}(u) = G_{IJKL}^{\dot \mu_1 \dot \mu_2 \dot \mu_3 \dot \mu_4}
\z_1^{\dot \mu_1} \z_2^{\dot \mu_2} \z_3^{\dot \mu_3} \z_4^{\dot \mu_4},
$$
where
\bea
\nonumber
&&
G_{IJKL}^{\dot \mu_1 \dot \mu_2 \dot \mu_3 \dot \mu_4} = 
\l \s_{g_\infty}[\k_3/2,\k_4/2](\infty) \t^i_{IJ}(1) \t^j_{KL}(u)
    \s_{g_0}[\k_1/2,\k_2/2](0) \r \,
\l \S_{g_\infty}^{\dot \mu_3 \dot \mu_4}(\infty) \S^i_{IJ}(1) \S^j_{KL}(u)
    \S_{g_0}^{\dot \mu_1 \dot \mu_2}(0) \r \\
\la{gu}
&& \equiv
\l \t_i\t_j \r (u) \, G_{IJKL}^{\dot \mu_1 \dot \mu_2 \dot \mu_3 \dot \mu_4 ij}(u).
\eea
\noindent Without any loss of generality, we will always assume that the
polarization $\z^{\dot \mu \nu}$ can be taken in the form
$\z^{\dot \mu} \z^{\nu}$.

In the approach of \cite{AF2}, the calculation of the correlation function
$G_{IJKL}(u)$ was based on the stress-energy tensor
method \cite{DFMS} which requires the knowledge of the Green function for
$DN$ bosonic fields $X^i_I(z)$, $I=1,\ldots,N$,$i=1,\ldots,D$. Recall that
$X^i_I(z)$
have cyclic boundary conditions (\ref{cyc}) around the insertion points of the
twist fields $\s_{(n)}(z)$ and therefore the corresponding Green function
is $N$-valued. So, to find the Green
function, and consequently the correlation function $G_{IJKL}$, one needs to
construct the $N$-fold map from the $z$-plane, on which
it is multi-valued, to the sphere, which we call the $t$-sphere, on which it is
single-valued. According to \cite{AF2} this map is unique, and is given by
the formula
\be
\la{map}
z = {\left(\frac{t}{t_1}\right)}^{n_0}
    {\left(\frac{t-t_0}{t_1-t_0}\right)}^{N-n_0}
    {\left(\frac{t_1-t_\infty}{t-t_\infty}\right)}^{N-n_\infty} \equiv u(t),
\ee
\noindent where we require the point $t=x$ to be mapped to $z=u$.
Due to the projective invariance, the positions of points
$t_0$, $t_1$, and $t_\infty$ can be chosen to depend on $x$ in a specific
manner,
that is $t_0 = t_0(x)$, $t_1 = t_1(x)$, and $t_\infty = t_\infty(x)$, and one
possible choice of this dependence is described in \cite{AF2}.
If we make the substitution (see \cite{AF2}):
\bea
\nonumber
&&
t_0 = x-1, \\
\nonumber
&&
t_\infty = x - \frac{(N-n_\infty)x}{(N-n_0)x + n_0} , \\
\nonumber
&&
t_1 = \frac{N-n_0-n_\infty}{n_\infty} + \frac{n_0 x}{n_\infty} - 
\frac{N(N-n_\infty)x}{n_\infty((N-n_0)x+n_0)}
\eea
eq.(\ref{map}) transforms into a function of $x$ alone which
can be viewed as the $2(N-n_0)$-fold covering of the $u$-sphere by the
$x$-sphere. Since the number of nontrivial correlation functions in
(\ref{fuu1}) is also equal to $2(N-n_0)$, as one can easily verify, we see
that the $t$-sphere can be represented as the union of $2(N-n_0)$ domains
and each domain, denoted by $V_{IJKL}$, contains the point $x$ corresponding
to some correlation function from (\ref{fuu1}).

Finally note that as was shown in \cite{AF2} the overall phase of
$G_{IJKL}(u)$ can not be determined and in principle
can depend on the indices $I,J,K,L$.
However, below we will show that the correlation function of the
holomorphic sector is complex-conjugated to the correlation function
of the anti-holomorphic sector. Therefore, by combining the two sectors 
the phase ambiguity disappears. To prove this assertion we have to take 
into account the symmetry of a correlation function under the change
$$
\s_{g}[\k/2] \leftrightarrow \bar{\s}_{g^{-1}}[{\tilde \k}/2] \quad {\rm and}
\quad
\S_{g}^{\dot \mu} \leftrightarrow \bar{\S}_{g^{-1}}^{\mu}
$$
to obtain the equality
\bea
\nonumber
&&
\l V_{g_\infty}[\k_3,\z_3;\k_4,\z_4](\infty)V_{IJ}(1)V_{KL}(u)
V_{g_0}[\k_1,\z_1;\k_2,\z_2](0)\r \\
\nonumber
&&=
\l \bar{V}_{g_{\infty}^{-1}}
[{\tilde \k}_3,{\tilde \z}_3;{\tilde \k}_4,{\tilde \z}_4](\infty)
\bar{V}_{IJ}(1) \bar{V}_{KL}(u)
\bar{V}_{g_0^{-1}}
[{\tilde \k}_1,{\tilde \z}_1;{\tilde \k}_2,{\tilde \z}_2](0)\r .
\eea
Then complex conjugating the obtained expression gives
\bea
\nonumber
&&
{\l \bar{V}_{g_{\infty}^{-1}}
[{\tilde \k}_3,{\tilde \z}_3;{\tilde \k}_4,{\tilde \z}_4](\infty)
\bar{V}_{IJ}(1) \bar{V}_{KL}(u)
\bar{V}_{g_0^{-1}}
[{\tilde \k}_1,{\tilde \z}_1;{\tilde \k}_2,{\tilde \z}_2](0)\r}^* \\
\nonumber
&&=
\lim_{z_{\infty} \to \infty} \lim_{z_0 \to 0}
z_{\infty}^{-2 \D_{g_\infty}[\k_3,\k_4]}
z_0^{-2 \D_{g_0}[\k_1,\k_2]} \,
u^{-3} \times \\
\nonumber
&&
\l \bar{V}_{g_0}
[-{\tilde \k}_1,{\tilde \z}_1;-{\tilde \k}_2,{\tilde \z}_2](\frac{1}{z_0})
\bar{V}_{KL}(\frac{1}{u}) \bar{V}_{IJ}(1) \bar{V}_{g_\infty}
[-{\tilde \k}_3,{\tilde \z}_3;-{\tilde \k}_4,{\tilde \z}_4](\frac{1}{z_\infty})\r.
\eea
Due to the $SO(8)$ invariance of the correlation function we can make the
replacement $-{\tilde \k} \to \k $, ${\tilde \z} \to \z $ and after performing
the conformal transformation $z \to \frac{1}{z}$ obtain
\bea
\nonumber
&&
{\l V_{g_\infty}[\k_3,\z_3;\k_4,\z_4](\infty)V_{IJ}(1)V_{KL}(u)
V_{g_0}[\k_1,\z_1;\k_2,\z_2](0)\r}^* \\
\nonumber
&& =
\l \bar{V}_{g_{\infty}}[\k_3,\z_3;\k_4,\z_4](\infty)
\bar{V}_{IJ}(1)
\bar{V}_{KL}(u)
\bar{V}_{g_0}
[\k_1,\z_1;\k_2,\z_2](0)\r .
\eea
By making the formal substitution $z \to \bar{z}$ we arrive at the
correlation function of the anti-holomorphic sector containing right-moving
fermions instead of left-moving ones. Thus, if the anti-holomorphic sector 
is obtained from the holomorphic one by the substitution: 
$z\to\bar{z}$, left-moving fermion $\to$ right-moving fermion, then the 
overall phase of $G_{IJKL}(u,\bar{u})$ is irrelevant. 

Now we present the solution for the correlation function $G_{IJKL}(u)$
that was found in \cite{AF2}. In particular,
$\l \t_i \t_j \r (u)$ is given by
\bea
&&
\la{tt}
\l\t_i\t_j\r (u) = -\delta^{ij}\frac{4 x(x-1)\7 \8 \9 }
{(n_0-n_{\infty})(x-\alpha_1)^2(x-\alpha_2)^2} + {\l\t_i\t_j\r}_k,
\\
\nonumber
&& 
{\l\t_i\t_j\r}_k = \\
\nonumber
&&
-\left(
\frac{\7}{n_0}k_1^i+\frac{\8}{n_\infty} k_3^i+\frac{1}{N-n_0}k_4^i
\right)
\left((x-1)k_1^j+x k_3^j
+\frac{n_0-n_{\infty}}{N-n_{\infty}}\9 k_4^j\right),
\nonumber
\eea
while the correlation function
$G_{IJKL}^{\dot \mu_1 \dot \mu_2 \dot \mu_3 \dot \mu_4 i j} (u)$
is equal to
\bea
\nonumber
&&
G_{IJKL}^{\dot \mu_1 \dot \mu_2 \dot \mu_3 \dot \mu_4 i j} (u) = \\
\nonumber
&&=
{\kappa}^{1/2} \frac{i R^4 }{2^6 (n_\infty-n_0)(N-n_0)}
\left(\frac{n_{\infty} n_0 (N-n_{\infty})}{(N-n_0)}\right)^{1/2}
\left(\frac{n_\infty-n_0}{N-n_0}\right)^{\frac{1}{4}k_1 k_3}
\frac{\9^3}{u^{3/2}(x-\al_1)^2(x-\al_2)^2} \\
\la{holcor}
&&\times
\left(\frac{x\8}{\9}\right)^{1+\frac{1}{4}k_1k_4}
\left(\frac{(x-1)\7}{\9}\right)^{1+\frac{1}{4}k_3k_4}
T_{IJKL}^{\dot \mu_1 \dot \mu_2 \dot \mu_3 \dot \mu_4 ij}(u).
\eea
\noindent Here $T_{IJKL}^{\dot \mu_1 \dot \mu_2 \dot \mu_3 \dot \mu_4 ij}(u)$
is defined in the $SU(4) \times U(1)$ basis according to
\bea
\la{T_IJKL}
&&
T_{IJKL}^{{\cal A}_1 {\cal A}_2 {\cal A}_3 {\cal A}_4 {\cal A}_5 {\cal A}_6}(u)
= \\
\nonumber
&&
C(g_0,g_\infty)
\times
\frac{ x^{d_0}{\6}^{d_1}{\7}^{d_2}{\8}^{d_3}{\9}^{d_4}}{
{((x-\alpha_1)(x-\alpha_2))}^{d_5}} ,
\eea
the coefficients $d_i$ are given by
\bea
\nonumber
d_0&=&
\p_1\p_4+\p_6\p_1+\p_6\p_4,
\qquad
d_1=
\p_6\p_3+\p_6\p_4+\p_3\p_4,
\\
\nonumber
d_2&=&
\p_1\p_2+\p_6\p_1+\p_6\p_2,
\qquad
d_3=
\p_6\p_2+\p_6\p_3+\p_2\p_3,
\\
\nonumber
d_4&=&
\p_6\p_1+\p_6\p_3+\p_1\p_3,
\qquad
d_5=-\p_5\p_6,
\\
\nonumber
d_6&=&
\p_1\p_5+\p_3\p_5+\p_1\p_3-\p_2\p_6
\eea
and
\bea
\la{12a}
|C(g_0,g_{\infty})| = \frac{n_0^{\p_1\p_5}n_\infty^{\p_3\p_5}
(N-n_\infty)^{\p_4\p_6}
(n_\infty-n_0)^{d_4-d_5} }
{(N-n_0)^{d_6}}.
\eea
A few comments are in order. First,
recall that $\kappa$ was introduced in Section 2.2 as a multiplicative factor
which compensates for the nonabelian nature of the orbifold. This constant is
equal to $2^3$ (for derivation see \cite{AF2}).

Secondly, computation of the fermionic correlation function 
$\l \S_{g_\infty}^{\dot \mu_3 \dot \mu_4}(\infty)
\S^i_{IJ}(1) \S^j_{KL}(u) \S_{g_0}^{\dot \mu_1 \dot \mu_2}(0) \r$ was 
done by bosonizing the fermions \cite{AF2} in the framework of
the $SU(4)\times U(1)$ formalism \cite{GSW} which is concisely presented in
Appendix B. Here we only note that in this formalism there is a one-to-one
correspondence between the $SU(4) \times U(1)$ index ${\cal A} 
\equiv \{ A, \bar{A} \}$ and the weight vector $\p$. Specifically, 
if ${\cal A}$ corresponds to ${\bf 8_v}$ then 
$\p^{A}(\p^{\bar{A}}) = {\bf e}^{A}(-{\bf e}^{A})$ and if it corresponds 
to ${\bf 8_c}$ then
$\p^{A}(\p^{\bar{A}}) = \q^{\dot A}(-\q^{\dot A})$, where
$\pm {\bf e}^{A}$ has components $\d^A_B$ and $\pm \q^{\dot A}$ is defined in 
eq.(\ref{weight}).

Before we go on to consider the scattering amplitude
let us point out the remarkable property of ${\l\t_i\t_j\r}_k$, namely
\be
\la{tau+}
{\l\t_+\t_{\mu}\r}_k = 0 = {\l\t_{\mu}\t_+\r}_k.
\ee
To prove this assertion, we note that the $"+"$ light-cone component 
of the first factor in ${\l \t_i \t_j \r}_k$ is equal to
\bea
&& 
\frac{\7}{n_0}k^+_1 + \frac{\8}{n_{\infty}}k^+_3 +\frac{1}{N-n_0}k^+_4 \\
\nonumber
&& =
\frac{1}{N} \left(
\7 - \8 - \frac{N-n_{\infty}}{N-n_0} \right) = 0,
\eea
while the $"+"$ light-cone component of the second factor is equal to
\bea
\nonumber
&& 
\6 k^+_1 + \5 k^+_3 + \frac{ n_0-n_{\infty} }{ N-n_{\infty} } \9 k^+_4 \\
\nonumber
&&  =
\frac{1}{N} \left(
n_0 \6 - n_{\infty} \5 - (n_0-n_{\infty})\9 \right) = 0.
\eea
This property will be used in establishing the Lorentz invariance of
scattering amplitudes.

\paragraph{Scattering amplitude}
Up to now we considered the correlation function
$G_{IJKL}^{\dot \mu_1 \dot \mu_2 \dot \mu_3 \dot \mu_4}(u,\bar{u})$
correspoding to $|u|<1$. It turns out that the correlation function 
corresponding to $|u|>1$ is again given by (\ref{holcor}) and so the 
time-ordering in (\ref{fuu}) can be omitted. Consequently, from 
(\ref{fuu1}),(\ref{guu}) and (\ref{gu}) we find that ${\cal M}$ is equal to
$$
{\cal M} = 2 N^2 \sqrt{k_1^+k_2^+k_3^+k_4^+}\sum_{IJKL}
\int d^2u |u| G_{IJKL}^{\dot \mu_1 \dot \mu_2 \dot \mu_3 \dot \mu_4}(u)
{\bar G}_{IJKL}^{\nu_1 \nu_2 \nu_3 \nu_4}(\bar u) \z_1^{\dot \mu_1 \nu_1}
\z_2^{\dot \mu_2 \nu_2} \z_3^{\dot \mu_3 \nu_3} \z_4^{\dot \mu_4 \nu_4}
$$
Substituting (\ref{holcor}) for the holomorphic part of the correlation
function  $G_{IJKL}(u,\bar u)$ and its complex conjugate for the 
anti-holomorphic part so as to get rid of the phase ambiguity, we arrive at 
the following expression for ${\cal M}$
\bea
\nonumber
&&
{\cal M} =
\frac{R^8}{2^{8} \sqrt{k_1^+k_2^+k_3^+k_4^+}}
{\left( \frac{n_0 n_\infty (N-n_\infty)}{N(N-n_0)}
\right)}^2 {\left(\frac{n_\infty-n_0}{N-n_0}\right)}^{\frac12 k_1 k_3} \\
\nonumber
&&\times
\int d^2u {\left| \frac{du}{dx} \right|}^2 
{\left| \frac{\5\8}{\9} \right|}^{\frac12 k_1k_4}
{\left| \frac{\6\7}{\9} \right|}^{\frac12 k_3k_4} \\
\nonumber
&&\times
\sum_{IJKL}
T_{IJKL}^{\dot \mu_1 \dot \mu_2 \dot \mu_3 \dot \mu_4}(u)
T_{IJKL}^{\nu_1 \nu_2 \nu_3 \nu_4}(\bar u) \z_1^{\dot \mu_1 \nu_1}
\z_2^{\dot \mu_2 \nu_2} \z_3^{\dot \mu_3 \nu_3} \z_4^{\dot \mu_4 \nu_4},
\eea
where we introduced a concise notation
\be
\la{concise}
T_{IJKL}^{\dot \mu_1 \dot \mu_2 \dot \mu_3 \dot \mu_4}(u) =
\l \t_i\t_j \r T_{IJKL}^{\dot \mu_1 \dot \mu_2 \dot \mu_3 \dot \mu_4ij}(u).
\ee
Recall that under the transformation $u \to x$
the $u$-sphere is mapped onto the domain $V_{IJKL}$. Taking this into account
and performing the change of variables \cite{AF}
\be
\la{z}
z = \frac{\5 \8}{\frac{n_\infty-n_0}{N-n_0}\9} \quad \Rightarrow \quad
dz = \frac{(x-\alpha_1)(x-\alpha_2)}{\frac{n_\infty-n_0}{N-n_0}{\9}^2}dx,
\ee
the expression for ${\cal M}$ assumes the conventional form
\bea
\la{M1}
&&
{\cal M} =
\frac{R^8}{2^8 \sqrt{k_1^+k_2^+k_3^+k_4^+}}
{\left( \frac{n_0 n_\infty (N-n_\infty)}{N(N-n_0)} \right) }^2 \\
\nonumber
&&\times
\int d^2z {\left| \frac{dx}{dz} \right|}^2
{| z |}^{\frac12 k_1k_4}
{| 1-z |}^{\frac12 k_3k_4}
T^{\dot \mu_1 \dot \mu_2 \dot \mu_3 \dot \mu_4}(z)
T^{\nu_1 \nu_2 \nu_3 \nu_4}(\bar{z}) \z_1^{\dot \mu_1 \nu_1}
\z_2^{\dot \mu_2 \nu_2} \z_3^{\dot \mu_3 \nu_3} \z_4^{\dot \mu_4 \nu_4}.
\eea
Now it follows from eq.(\ref{S}) that in the limit $R \to \infty$ the
expression for the $S$-matrix element to the second order in the coupling 
constant $\lambda$ is given by 
\bea
\la{S1}
&&
\l f| S | i\r = 
-i \lambda^2 2^{-8} 2 N \d_{m_1+m_2+m_3+m_4,0} \d (\sum_i k_i^-) \d^D(\sum_i \k_i)
\sqrt{ \frac{ 
\prod_{i=1}^{4} {(k_i^+)}^{\epsilon{\scriptscriptstyle{(\dot \mu_i)}}} 
{(k_i^+)}^{\epsilon{\scriptscriptstyle{(\nu_i)}}} }{
\prod_{i=1}^{4} k_i^+} } \,
{\cal I}(\z;k)
\eea
where 
\bea
\la{I}
&&
{\cal I}(\z;k) =
{\left(\frac{n_0 n_\infty (N-n_\infty)}{N-n_0}\right)}^2
\sqrt{ \prod_{i=1}^{4} {(k_i^+)}^{-\epsilon{\scriptscriptstyle{(\dot \mu_i)}}}
{(k_i^+)}^{-\epsilon{\scriptscriptstyle{(\nu_i)}}} } \\
\nonumber
&& \times
\int d^2z {\left| \frac{dx}{dz} \right|}^2
\left|z\right|^{\frac{1}{2}k_1k_4}
\left|1-z\right|^{\frac{1}{2}k_3k_4}
T^{\dot \mu_1 \dot \mu_2 \dot \mu_3 \dot \mu_4}(z)
T^{\nu_1 \nu_2 \nu_3 \nu_4}(\bar{z})
\z_1^{\dot \mu_1 \nu_1} \z_2^{\dot \mu_2 \nu_2} \z_3^{\dot \mu_3 \nu_3} 
\z_4^{\dot \mu_4 \nu_4}.
\eea
Here $\epsilon^{\scriptscriptstyle{(\dot \mu_i)}}(
\epsilon^{\scriptscriptstyle{(\nu_i)}})$
is equal to $0$ if $\dot \mu_i (\nu_i)$ corresponds to ${\bf 8_v}$ and
is equal to $1$ if $\dot \mu_i (\nu_i)$ corresponds to ${\bf 8_c}$ 
$({\bf 8_s})$. Also note that we have restored $\d$-functions responsible 
for the momentum conservation law and represented the light-cone momenta 
$k_i^+$ as $k_i^+=\frac {m_i}{N}$. In the next section we will compute
all open string kinematical factors and show that all dependence
on $N$ in ${\cal I}(\z;k)$ is absorbed into the light-cone momenta $k_i^+$ 
and hence we are justified to consider the limit $N \to \infty$ in 
(\ref{S1}). In this limit the combination  
$N \d_{m_1+m_2+m_3+m_4,0}$ goes to $\d(\sum_i k_i^+)$ and formula (\ref{S1})
acquires the form
\be
\la{S2}
\l f| S | i\r = 
-i \lambda^2 \d^{D+2}(\sum_i k_i^{\mu}) 
\sqrt{ \frac{ 
\prod_{i=1}^{4} {(k_i^+)}^{\epsilon{\scriptscriptstyle{(\dot \mu_i)}}} 
{(k_i^+)}^{\epsilon{\scriptscriptstyle{(\nu_i)}}} }{
\prod_{i=1}^{4} k_i^+} }
2^{-8} {\cal I}(\z;k).
\ee
In order to extract the scattering amplitude $A(1,2,3,4)$ from the 
$S$-matrix element one needs to make use of the reduction formula, namely 
\be
\la{amplitude}
\langle f|S|i\rangle =
-i \d^{D+2}(\sum_i \k_i^{\mu})
\sqrt{ \frac{ 
\prod_{i=1}^{4} {(k_i^+)}^{\epsilon{\scriptstyle{(\dot \mu_i)}}} 
{(k_i^+)}^{\epsilon{\scriptstyle{(\nu_i)}}} }{
\prod_{i=1}^{4} k_i^+} } A(1,2,3,4).
\ee
Using eq.(\ref{S2}) for the $S$-matrix element and taking into
account the reduction formula we obtain the general expression for
the four particle scattering amplitude $A(1,2,3,4)$:
$$
A(1,2,3,4) = \lambda^2 2^{-8} {\cal I}(\z;k).
$$

Consequently, the problem of finding $A(1,2,3,4)$ in the 
$S^N \R^8$ orbifold sigma model is reduced to the calculation of 
${\cal I}(\z;k)$. In the next Section we will find that ${\cal I}(\z;k)$ 
can be written in the form which is standard in the superstring theory, 
namely
\be
\la{master}
{\cal I}(\z;k) = K(\z;k) K(\z;k) C(s,t,u),
\ee
where
\be
\la{C_stu}
C(s,t,u) = - \pi \frac{\G(-s/8) \G(-t/8) \G(-u/8)}{\G(1+s/8) \G(1+t/8)
\G(1+u/8)}.
\ee
Here we introduced open string kinematical factors $K(\z;k)$ 
which we will show coincide with the well-known kinematical factors
obtained in the framework of the supersring theory.

\section{Kinematical factors}
\subsection{vector particle+vector particle $\to$ fermion+fermion}
To conform with the standard notation of the supersting theory let us denote 
the polarization of a left-(right-) moving fermion by $u^{\a} (u^a)$ instead 
of $\z^{\a} (\z^{a})$ preserving $\z$ for polarizations of massless 
vector particles.

As follows from eq.(\ref{master}) and (\ref{I}) in order to find the 
kinematical factor corresponiding to two massless vector particles in the 
inital state (i.e. $\dot \mu_1 \to i_1$, $\dot \mu_2 \to i_2$) and two 
fermions in the final state (i.e. $\dot \mu_3 \to \a_3$, $\dot \mu_4 \to 
\dot a_4$) one first has to find
\bea
\la{T1}
T^{i_1 i_2 \a_3 \a_4}(z) = \l \t_i \t_j \r(z) T^{i_1 i_2 \a_3 \a_4 i j}(z),
\eea
where the spin-tensor $T^{i_1 i_2 \a_3\a_4 ij}(z)$ is determined up to an 
unknown phase by (\ref{T_IJKL}). The overall phase is irrelevant
in our computations since we choose the kinematical factor of the right-moving
sector to coincide with that of the left-moving sector. Nevertheless it is essential to 
know the relative phases of $T^{i_1 i_2 \a_3\a_4 ij}(z)$ for different values 
of $SO(8)$ indices $i_m$ and $\a_n$. In order to fix these phases we decompose 
the spin-tensor $T^{i_1 i_2 \a_3\a_4 ij}(z)$ into the sum of $SO(8)$ 
invariant rank six spin-tensors:
\bea
\nonumber
&&T^{i_1i_2\a_3\a_4ij}(z)=\\
\nonumber
&&=\frac14 {\g^{[i_1i_2ij]}}_{\a_3\a_4} C_1(z)\\
\nonumber
&&
+\frac12 {\g^{[i_1i_2]}}_{\a_3\a_4} \d^{i j} C_2(z)
+\frac12 {\g^{[i_2j]}}_{\a_3\a_4} \d^{i_1 i} C_3(z)
+\frac12 {\g^{[i_2i]}}_{\a_3\a_4} \d^{i_1 j} C_4(z)\\
\nonumber
&&+\frac12 {\g^{[i_1j]}}_{\a_3\a_4} \d^{i_2 i} C_5(z)
+\frac12 {\g^{[i_1i]}}_{\a_3\a_4} \d^{i_2 j} C_6(z)
+\frac12 {\g^{[ij]}}_{\a_3\a_4} \d^{i_1i_2} C_7(z)\\
\nonumber
&&+\,\,\,\,\,\d_{\a_3\a_4} \d^{i_1i_2} \d^{ij} C_8(z)
  +\,\,\,\,\,\d_{\a_3\a_4} \d^{i_1i} \d^{i_2j} C_9(z)
  +\,\,\,\,\,\d_{\a_3\a_4} \d^{i_1j} \d^{i_2i} C_{10}(z).
\eea
By using the $SU(4)\times U(1)$ basis, the function $C_1(z)$
and $C_2(z)$ can be determined up to a phase using the following relations
\bea
\nonumber
&& T^{\bar 1 2 \dot 4 \dot 4 \bar 3 \bar 4} = -C_1 ,\\
\nonumber
&& T^{\bar 1 \bar 4 \dot 3 \dot 4 \bar 2 2} = -\frac12C_1 - C_2.
\eea
$$
\,\,\,\,\,\,\,\,T^{\bar 1 \bar 4 \dot 3 \dot 4 2 \bar 2} =  \frac12C_1 - C_2.
$$
Since we know all three functions up to a phase we get a
nontrivial equation on $C_1(z)$ and $C_2(z)$ allowing us to
determine their relative sign. Namely, from (\ref{T_IJKL}) with the account
of the normalization constant (\ref{12a}) one obtains
\bea
&&
\quad \quad \quad -C_1 \sim 
\frac{N-n_0}{n_{\infty}-n_0} \frac{e^{i\varphi_1}}{\5\8\9},
\\
\nonumber
&&
-\frac12 C_1 - C_2 \sim \frac{N-n_0}{n_{\infty}-n_0}
\frac{e^{i\varphi_2}}{\5(x-\alpha_1)(x-\alpha_2)},
\\
\nonumber
&&
\,\,\,\,\,\frac12 C_1 - C_2 \sim \frac{n_{\infty}(N-n_{\infty})}
{(n_{\infty}-n_0)^2} \frac{e^{i\varphi_3}}{\8\9(x-\alpha_1)(x-\alpha_2)},
\eea
where a common multiplier in all three functions was omitted. Now it can be 
easily verified that the last equation is satisfied only if
$e^{i\varphi_1} = e^{i\varphi_2} = -e^{i\varphi_3}$. Since we proved that
the overall phase is irrelevant, we can set $e^{i\varphi}=1$ and proceeding 
in the same manner fix relative signs of all 10 functions $C_i(z)$. For 
convenience in later computations it is useful to rewrite
$T^{i_1i_2\a_3\a_4ij}(z)$ in terms of ordinary products of $\g$'s
instead of their antisymmetric combinations. This is achieved
with the help of identity (\ref{identity1'}).
The final answer for $T^{i_1i_2i_3i_4ij}(z)$ is
\bea
\la{nonumber}
&&T^{i_1 i_2 \a_3 \a_4 i j}(z)= {(n_\infty(N-n_\infty))}^{-\frac12} \\
\nonumber
&&\times \left\{\,\,
 \frac14 \frac{N-n_0}{n_{\infty}-n_0}
\frac{{(\g^{i_1} \g^{i_2} \g^i \g^j)}_{\a_3 \a_4}}{\5\8\9} \right.\\
\nonumber
&&\,\,\,\,\,\,\,\,
+ \frac12
\frac{(\g^i \g^j)_{\a_3 \a_4} \d^{i_1 i_2}}{\5\6\7\8} \\
\nonumber
&&\,\,\,\,\,\,\,\,
- \frac12 \frac{N-n_0}{n_0}
\frac{{(\g^{i_2} \g^j)}_{\a_3 \a_4} \d^{i_1 i} }{\5\6\8} \\
\nonumber
&&\,\,\,\,\,\,\,\,
- \frac12 \frac{N-n_{\infty}}{n_{\infty}-n_0}
\frac{{(\g^{i_2} \g^i)}_{\a_3 a_4} \d^{i_1 j}}{\5\7\8\9} \\
\nonumber
&&\,\,\,\,\,\,\,\,
- \frac12 \frac{N-n_0}{n_{\infty}-n_0}
\frac{{(\g^{i_1} \g^j)}_{\a_3 \a_4} \d^{i_2 i}}{\5\6\9} \\
\nonumber
&&\,\,\,\,\,\,\,\,
+ \frac12 \frac{N-n_{\infty}}{n_{\infty}-n_0}
\frac{{(\g^{i_1} \g^i)}_{\a_3 \a_4} \d^{i_2 j}}{\7\8\9} \\
\nonumber
&&\,\,\,\,\,\,\,\,
- \frac12 \frac{n_{\infty}(N-n_{\infty})}{(n_{\infty}-n_0)^2}
\frac{{(\g^{i_1} \g^{\i_2})}_{\a_3 \a_4} \d^{ij}}{(x-\alpha_1)(x-\alpha_2)\8\9}\\
\nonumber
&&\,\,\,\,\,\,\,\,
+ \frac{N-n_{\infty}}{n_{\infty}-n_0}
\frac{\d^{i_1 j} \d^{i_2 i} \d_{\a_3 \a_4}}{\5\6\7\9} \\
\nonumber
&&\,\,\,\,\,\,\,\,
+ \frac{N-n_{\infty}}{n_0}
\frac{ \d^{i_1 i} \d^{i_2 j} \d_{\a_3 \a_4}}{\6\7\8} \\
\nonumber
&&\,\,\,\,\,\,\,\,
-\left.
\frac{n_{\infty}(N-n_{\infty})}{(n_{\infty}-n_0)(N-n_0)}
\frac{\d^{i_1i_2} \d^{ij} \d_{\a_3 \a_4}}{(x-\alpha_1)(x-\alpha_2)\6\7\8}\,\,
\right\}.
\eea
Next we contract $T^{i_1i_2\a_3\a_4ij}(z)$ with $\l \t_i \t_j \r$ and
substitute the result thus obtained into (\ref{T1}). 
After long and tedious calculations we arrive at the following expression for
${\cal I}(\z;k)$:
\bea
\la{int3}
&&
{\cal I}(\z;k) =
\int d^2 z {\left|z\right|}^{\frac12 k_1k_4 - 2}
{\left| 1-z \right|}^{\frac12 k_3k_4 - 2}  
T[u_3,\z_2,\z_1,u_4](z) T[u_3,\z_2,\z_1,u_4](\bar z) ,
\eea
where 
\bea
\la{K}
&&
T[u_3,\z_2,\z_1,u_4](z) = \frac{N}{4 n_\infty} \, \biggl\{
 (z-1){(\g^{i_1}\g^{i_2}\g^i\g^j)}_{\a_3\a_4} t^{ij} \\
\nonumber
&& 
+2{(\g^i\g^j)}_{\a_3\a_4} \d^{i_1i_2} t^{ij}
-2{(\g^{i_1}\g^i)}_{\a_3\a_4} p^{i i_2}
-2{(\g^{i_2}\g^i)}_{\a_3\a_4} q^{i i_1}
-4\d_{\a_3\a_4}\rho^{i_1 i_2} \biggr\} 
\z_1^{i_1} \z_2^{i_2} u_3^{\a_3} u_4^{\a_4}.
\eea
Here to simplify the notation we introduced the following tensors
\bea
\nonumber
&&
t^{ij}= k^i_3k^j_1 + \frac{n_\infty}{N-n_\infty} k^i_1k^j_4 +
\frac{n_0}{N-n_\infty} k^i_3k^j_4+\frac{n_\infty(N-n_\infty)}{n_0}k^i_1k^j_1,\\
\nonumber
&&
p^{i i_2}=\frac{n_0 n_\infty}{n_{\infty}-n_0} \left(
\frac{ N-n_{\infty} }{ n_{\infty}-n_0 }
\frac{\5\6}{\9}
\frac{ {\l\t_{i_2}\t_i\r}_k - {\l\t_i\t_{i_2}\r}_k }{(x-\alpha_1)(x-\alpha_2)}
+\frac{ {\l\t_{i_2}\t_i\r}_k }{\9} \right), \\
\nonumber
&&
q^{i i_1} = \frac{n_\infty (N-n_0)}{n_{\infty}-n_0} \left(
\7 \frac{{\l\t_{i_1}\t_i\r}_k - {\l\t_i\t_{i_1}\r}_k}{(x-\alpha_1)(x-\alpha_2)} +
\frac{{\l\t_i\t_{i_1}\r}_k}{\9} \right), \\
\la{tensors}
&&
\rho^{i_1 i_2} = \frac{n_\infty(N-n_{\infty})}{n_{\infty}-n_0} \left(
\5 \frac{{\l\t_{i_1}\t_{i_2}\r}_k - {\l\t_{i_2}\t_{i_1}\r}_k}{(x-\alpha_1)(x-\alpha_2)} +
\frac{{\l\t_{i_2}\t_{i_1}\r}_k}{\9} \right).
\eea

Note that we purposefully wrote these tensors in terms of the variable
$x$ even though it presents no difficulty to express them in terms of $z$.
The point is that by writing them in this form we can clearly see
that all $"+"$ light-cone components vanish due to (\ref{tau+}).

Next we turn to the issue of the Lorentz invariance of the
theory. To this
end, we introduce ten pure imaginary 32$\times$32 $\G$-matrices which satisfy
the Clifford algebra $\{ \G^{\mu},\G^{\nu} \} = -2\eta^{\mu\nu}$.  These
$\G$ matrices are constructed as tensor products of 2$\times$2 Pauli matrices
$\s_i, i=1,2,3$ and 16$\times$16 matrices $\g^i$ $i=1,\ldots,8$:
$$
\g^i = \left( \begin{array}{cc}0&\g^i_{a\a}\\ \g^i_{\a a}&0 \end{array}\right),
$$
where $\g^i_{a\a}$ and $\g^i_{\a a}$ are defined in (\ref{G}).
Light-cone components of ten-dimensional gamma matrices, i.e. 
$\G^+ = \G^0 + \G^9$ and $\G^- = \G^0-\G^9$, are nilpotent: 
${(\G^+)}^2 = {(\G^-)}^2 = 0$. Evidently, in the integrand (\ref{int3}) 
transversal components of ten-dimensional matrices will be contracted with 
fermion wave functions $u^{\a}$ $(u^a)$. In the light-cone coordinates 
the 32-component Majorana-Weyl spinor $u$, $\G_{11}u=+u$, assumes the form 
$(u^{\a},0,0,u^a)$. This spinor satisfies the massless Dirac equation
$k_\mu \G^{\mu} u=0$, or equivalently ${\bar u} \G^{\mu} k_{\mu} =0$, where
${\bar u}$ is the Dirac conjugated spinor, i.e. ${\bar u} = u^{\rm T} \G^0$.
In the chosen basis the Dirac equation takes the form (see e.g. \cite{GSW})
\bea
\la{dirac1}
&&k^+ u^a + \g^i_{a\a} k^i u^{\a} = 0, \\
\la{dirac2}
&&k^- u^{\a} + \g^i_{\a a} k^i u^a = 0.
\eea
The first of these equations allows one to express $u^a$ in terms of
$u^{\a}$:
\be
\la{ua}
u^a=-\frac{1}{k^+}\g^i_{a\a}k^iu^{\a} .
\ee
Therefore, eight components of $u^{\a}$ correspond to eight physical
degrees of freedom. Upon the substitution of (\ref{ua}) into eq.(\ref{dirac2})
one obtains the equation on $u^{\a}$ which is just the Klein-Gordon equation
$k^2=0$.
In order to express the integrand (\ref{K}) in terms of
ten-dimensional $\G$-matrices and 32-component Majorana-Weyl spinors
$u_i$ we need the following identities:
$$
\begin{array}{c}
u_1^{\a} {(\g^{i_1} \g^{i_2} \g^{i_3} \g^{i_4})}_{\a \b} u_2^{\b} =
\frac12 {\bar u}_1 \G^+ \G^{i_1} \G^{i_2} \G^{i_3} \G^{i_4} u_2, \\
\,\,\,\,u_1^{\a} {(\g^{i_1} \g^{i_2})}_{\a \b} u_2^{\b} =
-\frac12 {\bar u}_1 \G^+ \G^{i_1} \G^{i_2} u_2, \\
u_1^{\a} \d^{\a \b} u_2^{\b} =
\frac12 {\bar u}_1 \G^+ u_2,
\end{array}
$$
which can be easily verified by using the explicit form of
$\G$-matrices, provided in Appendix A. Now it is straightforward to
replace transversal 8$\times$8 $\g$-matrices with 32$\times$32 $\G$-matrices
and 8-component spinors $u^{\a}$, $u^a$ with 32-component Majorana-Weyl spinors
$u$.
In addition to fermion wave functions, the integrand (\ref{int3}) also
depends on vector polarizations. As usual, in ten dimensions a polarization of
a massless vector particle
satisfies the transversality condition: $k_{\mu}\z^{\mu}=0$.
In the light-cone gauge the polarization obeys $\z^+=0$
allowing us to express the component $\z^-$ in terms of $\z^i$ and $k_{\mu}$ as
$\z^-=\frac{2k^i\z^i}{k^+}$. In our model we only deal with eight transversal
polarizations $\z^i$ and can treat this equation as the definition of the
light-cone polarization $\z^-$.
An important property of the light-cone gauge is that
$\z_1^{i}\z_2^{i}=\z_1^{\mu}\z_2^{\mu}\equiv (\z_1\z_2)$ which is a direct
consequence of $\z^+_1 =\z^+_2=0$.
Clearly, the integrand in (\ref{int3}) depends on scalar products of
transversal momenta $k^i$ with $\z^i$.  It turns out that by using the
light-cone momenta and polarizations $k^-$ and $\z^-$ the integrand can be
written via scalar products of ten-dimensional vectors. To show that this
is indeed the case, we first note that $t^{i+} = t^{+i} = 0$ and the
same holds for all tensors in (\ref{tensors}). This is
a direct consequence of (\ref{tau+}). Taking into account $\{ \G^i,\G^+ \}=0$ 
and ${(\G^+)}^2=0$ the first term in (\ref{K}) becomes
\bea
\nonumber
&&
{\bar u}_1 \G^+ \G^{i_1}\z^{i_1} \G^{i_2}\z^{i_2}
\G^it^{ij}\G^j u_2 = \\
\nonumber
&&=
{\bar u}_1 \G^+ \G^{i_1}\z^{i_1} \G^{i_2}\z^{i_2} \G^i
(\frac12 \G^+t^{i-} + t^{i\nu}\G^{\nu}) u_2
=
{\bar u}_1 \G^+ \G^{i_1}\z^{i_1} \G^{i_2}\z^{i_2}
\G^i t^{i\nu}\G^{\nu} u_2 =\\
\nonumber
&&= \ldots
=
{\bar u}_1 \G^+ \G\z_1 \G\z_2 \G^{\mu}t^{\mu\nu}\G^{\nu} u_2.
\eea
Proceeding in the same manner we find 
$$
\begin{array}{c}
\,\,\,\,\,\,\,\,\,\,{\bar u}_1 \G^+\G^i\l\t_i\t_j\r\G^j u_2 =
{\bar u}_1 \G^+ \G^{\mu}\l\t_{\mu}\t_{\nu}\r\G^{\nu} u_2, \\
{\bar u}_1 \G^+ \G^{i_1}\z^{i_1} \G^{i_2}\z^{i_2} u_2 =
{\bar u}_1\G^+ \G \z_1 \G \z_2 u_2, \,\,\,\,\,\, \\
{\bar u}_1 \G^+ \G^{i_1}\z^{i_1} \G^i \l\t_i\t_{i_2}\r\z^{i_2} u_2 =
{\bar u}_1 \G^+ \G\z_1 \G^{\mu} \l\t_{\mu}\t_{\nu}\r\z^{\nu} u_2.
\end{array}
$$
Imposing the Dirac equation ${k_4}_{\mu} \G^{\mu} u_4=0$ and the transversality
condition ${k_1}_{\mu}{\z_1}^{\mu} =0= {k_2}_{\mu}{\z_2}^{\mu}$, the
expression for $T[u_3,\z_2,\z_1,u_4](z)$ acquires a particularly simple form:
\bea
\nonumber
&&
T[u_3,\z_2,\z_1,u_4](z) = \frac{N}{4 n_\infty} \,\biggl\{
(z-1) \frac12 {\bar u}_3 \G^+ \G\z_1 \G\z_2 \G k_3 \G k_1 u_4
- {\bar u}_3 \G^+ \G k_3 \G k_1  u_4 \z_1\z_2 \\
\nonumber
&&
+ 2 {\bar u}_3 \G^+ \G \z_1 [(z-1)\G k_1  \z_2 k_3 -
 z\,\G k_3 \z_2 k_1] u_4
+ 2 {\bar u}_3 \G\z_2 [(z-1)(\G k_3 \z_1 k_4 - \G k_1 \z_1 k_3)
+ 4 z\,\G k_3 \z_1 k_2] u_4 \\
\nonumber
&& 
- 2 {\bar u}_3 \G^+ u_4 [\z_1k_4 \z_2k_3 -z\,\z_1k_3
\z_2k_4] \biggr\}.
\eea

The last step in rendering $T[u_3,\z_2,\z_1,u_4](z)$ the Lorentz
covariant form, requires us to impose the Dirac equation
${\bar u}_3 \G^{\mu}{k_3}_{\mu} = 0$.
To this end, one has to anticommute $\G k_3$
all the way to the left until it multiplies the spinor
${\bar u}_3$ and annihilates it. This procedure will generate
additional terms due to
the anticommutation relation of $\G$-matrices. The appearance of these
terms can be easily traced in the example below:
\bea
\la{covar}
&&
\frac12 {\bar u}_3  \G^+   \G \z_1   \G \z_2   \G k_3
\G k_1 u_4 = -\frac12 {\bar u}_3 \G^+ \G \z_1 \G k_3 \G \z_2
\G k_1 u_4 - {\bar u}_3 \G^+ \G \z_1 k_3 \z_2 \G k_1 u_4 \\
\nonumber
&&
=\frac12 {\bar u}_3 \G^+ \G k_3 \G \z_1 \G \z_2 \G k_1 u_4
+ {\bar u}_3 \G^+ \z_1 k_3 \G\z_2 \G k_1 u_4
- {\bar u}_3 \G^+ \G\z_1 k_3\z_2 \G k_1 u_4 \\
\nonumber
&&
=-{\bar u}_3 \G\z_1 \G\z_2 \G k_1 u_4 k^+_3
+ {\bar u}_3 \G^+ \z_1 k_3 \G\z_2 \G k_1 u_4
- {\bar u}_3 \G^+ \G\z_1 k_3\z_2 \G k_1 u_4.
\eea
Proceeding in this fashion it can be easily shown that all terms
containing $\G^+$ cancel and we arrive at the following result:
\bea
\nonumber
&&
T[u_3,\z_2,\z_1,u_4](z) = 
-\frac{1-z}{4} {\bar u}_3 \G\z_2 \G(k_1+k_4) \G\ \z_1 u_4 \\
\nonumber
&& \qquad + \frac{z}{2} \left(
{\bar u}_3 \G \z_1 u_4  k_1 \z_2 - {\bar u}_3 \G \z_2 u_4  k_2 \z_1 -
{\bar u}_3 \G k_1 u_4  \z_1 \z_2 \right).
\eea
Finally, we perform the integration over the sphere $(z,\bar z)$ to get:
$$
{\cal I}(\z;k) = 
K(u_3,\z_2,\z_1,u_4;k) K(u_3,\z_2,\z_1,u_4;k) C(s,t,u),
$$
where
\bea
\la{K1}
&&
K(u_3,\z_2,\z_1,u_4;k) = 2^{-4} \,\biggl\{
-\frac{s}{2} {\bar u}_3 \G\z_2 \G(k_1+k_4) \G\ \z_1 u_4  \\
\nonumber
&&  
\qquad + t \left(
{\bar u}_3 \G \z_1 u_4  k_1 \z_2 - {\bar u}_3 \G \z_2 u_4  k_2 \z_1 -
{\bar u}_3 \G k_1 u_4  \z_1 \z_2 \right) \biggr\}.
\eea
Now one can recognize in $K(u_3,\z_2,\z_1,u_4;k)$ the standard open string 
kinematical factor of the superstring theory (see \cite{S}). Futhermore, as
was mentioned earlier all dependence on $N$ in
$K(u_1,\z_2,u_3,\z_4;k)$ was absorbed into $k^+$. 

\subsection{fermion+vector particle $\to$ fermion+vector particle}
The kinematical factor corresponding to a massless vector particle and a 
fermion in the initial state and the same type of particles in the final 
state is computed in complete analogy with the kinematical factor found in the
previous section. In  particular, here we need to determine the spin-tensor
$T^{\a_1 i_2 \a_3 i_4 ij}(z)$ which we decompose into $SO(8)$ invariant
rank six spin-tensors as follows
\bea
\nonumber
&&
T^{\a_1i_2\a_3i_4ij}(z)=\\
\nonumber
&&=
\frac14 {\g^{[i_2i_4ij]}}_{\a_1\a_3} C_1(z)\\
\nonumber
&&
+\frac12 {\g^{[ij]}}_{\a_1\a_3} \d^{i_2i_4} C_2(z)
+\frac12 {\g^{[i_2j]}}_{\a_1\a_3} \d^{i_4 i} C_3(z)
+\frac12 {\g^{[i_2i]}}_{\a_1\a_3} \d^{i_4 j} C_4(z)\\
\nonumber
&&
+\frac12 {\g^{[i i_4]}}_{\a_1\a_3} \d^{i_2 i} C_5(z)
+\frac12 {\g^{[j i_4]}}_{\a_1\a_3} \d^{i_2 j} C_6(z)
+\frac12 {\g^{[i_2i_4]}}_{\a_1\a_3} \d^{i j} C_7(z)\\
\nonumber
&&
+\,\,\,\,\,\d_{\a_1\a_3} \d^{i_2i_4} \d^{ij} C_8(z)
+\,\,\,\,\,\d_{\a_1\a_3} \d^{i_2i} \d^{i_4j} C_9(z)
+\,\,\,\,\,\d_{\a_1\a_3} \d^{i_2j} \d^{i_4i} C_{10}(z).
\eea
To fix the functions $C_i(z)$ we transform to the
$SU(4)\times U(1)$ basis, as we did in the previous case. After fixing
relative signs of $C_i(z)$ we arrive at the following expression
for $T^{\a_1i_2\a_3i_4ij}(z)$
\bea
\nonumber
&&
T^{\a_1i_2\a_3i_4 ij}(z) = {(n_0 n_\infty)}^{-\frac12} \\
\nonumber
&&\times \left\{\,\,
\frac14 \frac{{(\g^{i_1}\g^{i_2}\g^i\g^j)}_{\a_1 \a_3}}{\5\6\7\8} \right. \\
\nonumber
&&\,\,\,\,\,\,\,\,
+\frac12 \frac{N-n_0}{n_{\infty}-n_0} \
\frac{{(\g^i\g^j)}_{\a_1\a_3} \d^{i_2i_4}}{\5\8\9} \\
\nonumber
&&\,\,\,\,\,\,\,\,
-\frac12 \frac{n_{\infty}}{n_{\infty}-n_0}
\frac{{(\g^j \g^{i_4})}_{\a_1\a_3} \d^{i_2i}}{\5\6\8\9} \\
\nonumber
&&\,\,\,\,\,\,\,\,
-\frac12 \frac{{(\g^i \g^{i_4})}_{\a_1\a_3} \d^{i_2j}}{\5\7\8} \\
\nonumber
&&\,\,\,\,\,\,\,\,
-\frac12 \frac{n_{\infty}}{n_{\infty}-n_0}
\frac{{(\g^{i_2}\g^j)}_{\a_1\a_3} \d^{i_4i}}{\6\7\8\9} \\
\nonumber
&&\,\,\,\,\,\,\,\,
-\frac12 \frac{N-n_0}{N-n_{\infty}}
\frac{{(\g^{i_2}\g^i)}_{\a_1\a_3} \d^{i_4j}}{\5\6\7} \\
\nonumber
&&\,\,\,\,\,\,\,\,
+\frac12 \frac{n_0}{n_{\infty}-n_0}
\frac{{(\g^{i_2}\g^{\i_4})}_{\a_1\a_3} \d^{ij}}{\5\7(x-\alpha_1)(x-\alpha_2)} \\
\nonumber
&&\,\,\,\,\,\,\,\,
-\frac{n_0}{n_{\infty}-n_0}
\frac{\d^{i_2j}\d^{i_4i}\d_{\a_1\a_3} \6}{\5\7\8\9} \\
\nonumber
&&\,\,\,\,\,\,\,\,
+\frac{n_{\infty}(N-n_0)}{(n_{\infty}-n_0)(N-n_{\infty})}
\frac{\d^{i_2i}\d^{i_4j}\d_{\a_1\a_3} \7}{\5\6\8\9} \\
\nonumber
&&\,\,\,\,\,\,\,\, \left.
+\frac{n_0(N-n_0)}{(n_{\infty}-n_0)^2}
\frac{\d^{i_2i_4}\d^{ij}\d_{\a_1\a_3} \6}{\5\9 (x-\alpha_1)(x-\alpha_2)}
\right\}.
\eea

Next we contract 
$T^{\a_1 i_2 \a_3 i_4 i j}(z)$ 
with 
$\l \t_i \t_j \r(z)$ 
in order to obtain 
$T^{\a_1 i_2 \a_3 i_4} = T^{\a_1 i_2 \a_3 i_4 ij} \l \t_i \t_j \r$.
After long calculations we find that ${\cal I}(\z;k)$ is equal to
\bea
\nonumber
&&
{\cal I}(\z;k) =
\int d^2 z {\left|z\right|}^{\frac12 k_1k_4 - 2}
{\left| 1-z \right|}^{\frac12 k_3k_4 - 2} 
T^{\a_1 i_2 \a_3 i_4}(z) 
T^{a_1 j_2 a_3 j_4}(\bar z) u_1^{\a_1 a_1} 
\z_2^{i_2 j_2} u_3^{\a_3 a_3} \z_4^{i_4 j_4},
\eea
where
\bea
\nonumber
&&
T^{\a_1 i_2 \a_3 i_4}(z) = \\
\nonumber
&&=
\frac18 {(\g^{i_2} \g^{i_4} \g^{i} \g^{j})}_{\a_1 \a_3}
\frac{N(N-n_\infty)}{n_\infty-n_0} 
\frac{{\l \t_i \t_j \r}_k - {\l \t_j \t_i \r}_k}{
(x-\alpha_1)(x-\alpha_2)} \\
\nonumber
&&+
\frac14 {(\g^i \g^j)}_{\a_1 \a_3} \d^{i_2 i_4} 
\frac{N(N-n_\infty)}{n_\infty - n_0}
\frac{{\l \t_i \t_j \r}_k - {\l \t_j \t_i \r}_k}{
(x-\alpha_1)(x-\alpha_2)} 
(z-1) \\
\nonumber
&&-
\frac12 {(\g^{i_2} \g^{i})}_{\a_1 \a_3}   
\frac{N(N-n_0)}{n_\infty - n_0} \left( \frac{
{\l \t_{i} \t_{i_4} \r}_k - {\l \t_{i_4} \t_{i} \r}_k}{
(x-\alpha_1)(x-\alpha_2)}\,\8 + \frac{ {\l \t_{i_4} \t_i \r}_k}{\9} \right) \\ 
\nonumber
&&-
\frac12 {(\g^{i} \g^{i_4})}_{\a_1 \a_3}   
\frac{N(N-n_\infty)}{(n_\infty-n_0) \9}
\left( -\frac{n_\infty}{n_\infty-n_0} \frac{{\l \t_{i} \t_{i_2} \r}_k - 
{\l \t_{i_2} \t_{i} \r}_k}{(x-\alpha_1)(x-\alpha_2)} \, \7 +
{\l \t_{i} \t_{i_2} \r}_k \right) \\
\nonumber
&&+
\d_{\a_1 \a_3}   
\frac{N(N-n_\infty)}{(n_\infty-n_0) \9} \left( \frac{n_0}{n_\infty-n_0}
\frac{{\l \t_{i_2} \t_{i_4} \r}_k - {\l \t_{i_4} \t_{i_2} \r}_k}{
(x-\alpha_1)(x-\alpha_2)}  + \frac{N}{N-n_\infty} 
{\l \t_{i_2} \t_{i_4} \r}_k \right) \\
\nonumber
&&+
\frac14 {(\g^{i_2} \g^{i_4})}_{\a_1 \a_3} 
\frac{N}{n_0 n_\infty} \left( (N-n_0-n_\infty)k_3 k_4 + N k_1 k_4 \right) \\
\nonumber
&&
-\frac12 \d^{i_2 i_4} \d_{\a_1 \a_3}   
\frac{N}{n_0 n_\infty} \left((n_0 + n_\infty)z + (N-n_0-n_\infty) \right)
k_3 k_4
\eea
Note that in the last two lines we took advantage of (\ref{tau+}) in order
to obtain Lorentz invariant scalar products. To rewrite this expression 
in terms of ten dimentional $\G$-matrices and 32-component Majorana-Weyl
spinors $u_1$ and $u_3$ we should proceed exactly as we did in the previous
calculation. Namely, here we need the formulas
$$
\begin{array}{c}
u_1^{\a} {(\g^{i} \g^{j} \g^{k} \g^{l})}_{\a \b} u_3^{\b} =
\frac12 {\bar u}_1 \G^+ \G^{i} \G^{j} \G^{k} \G^{l} u_3, \\
\,\,\,\,u_1^{\a} {(\g^{i} \g^{j})}_{\a \b} u_3^{\b} =
-\frac12 {\bar u}_1 \G^+ \G^{i} \G^{j} u_3, \\
u_1^{\a} \d^{\a \b} u_3^{\b} =
\frac12 {\bar u}_1 \G^+ u_3.
\end{array}
$$
Taking into account these formulas as well as the property (\ref{tau+}),
the nilpotency of $\G^+$ and the fact that $\{ \G^i , \G^+ \} = 0$ then
after some algebra we find that $T[u_1,\z_2,u_3,\z_4] =
T^{\a_1 i_2 \a_3 i_4} u_1^{\a_1} \z_2^{i_2} u_3^{\a_3} \z_4^{i_4}$
is equal to
\bea
\nonumber
&&
T[u_1,\z_2,u_3,\z_4](z)
 = \frac{N}{4n_0} \, \biggl\{
\frac12 {\bar u}_1 \G^+\G\z_2 \G\z_4 \G k_1 \G k_4 u_3
- {\bar u}_1 \G^+ \G k_1 \G k_4 u_3 \z_2\z_4  \\
\nonumber
&&
+  {\bar u}_1 \G^+ (\G\z_2 \G^{\mu}\rho^{\mu\nu}\z_4^{\nu}
-  \G \z_4 \G^{\mu}p_{I}^{\mu \nu} \z_2^{\nu}) u_3
+ 2 {\bar u}_1 \G^+ u_3 (\z_2^{\mu} q^{\mu \nu} \z_4^{\nu} -
\z_4^{\mu} p_{II}^{\mu \nu} \z_2^{\nu}) \biggr\}.
\eea
Here for convenience we introduced the following tensors
\bea
\nonumber
&&
\rho^{\mu \nu} =  k_4^{\mu}k_1^{\nu} - zk_1^{\mu}k_3^{\nu}
-(z-1)k_1^{\mu}k_1^{\nu}, \\
\nonumber
&&
p^{\mu \nu}_{I} = (z-1)k_1^{\mu} k_4^{\nu} + k_4^{\mu} k_1^{\nu}, \\
\nonumber
&&
p^{\mu \nu}_{II} =(1-\frac{N}{n_\infty})k_3^{\mu}k_1^{\nu}
+ \frac{n_0}{n_\infty} (z-1)k_3^{\mu} k_4^{\nu} + (z-1)k_1^{\mu}k_4^{\nu}, \\
\nonumber
&&
q^{\mu \nu} = (z-1) k_1^{\mu} k_2^{\nu} + \frac{n_0}{n_\infty}(z-1)
k_4^{\mu}k_3^{\nu} -\frac{N}{n_\infty} k_1^{\mu}k_3^{\nu}.
\eea
To cast the integrand into the Lorentz covariant form we impose
the Dirac equation ${\bar u}_1 \G^{\mu} {k_1}_{\mu} = 0$.
Then all non-covariant terms, i.e. terms containing $\G^+$, cancel
and we obtain:
$$
{\cal I}(\z;k) = \int d^2 z {\left|z\right|}^{\frac12 k_1k_4 - 2}
{\left| 1-z \right|}^{\frac12 k_3k_4 - 2} 
T[u_1,\z_2,u_3\z_4](z)
T[u_1,\z_2,u_3\z_4](\bar z),
$$
where
\bea
\nonumber
&&
T[u_1,\z_2,u_3\z_4](z) = 
\frac{z}{4} {\bar u}_1 \G \z_2 \G(k_3+k_4) \G \z_4 u_3 \\
\nonumber
&&
+\frac{1-z}{4}{\bar u}_1 \G \z_4 \G(k_2+k_3) \G \z_2 u_3 .
\eea
Finally, we perform the integration over the sphere $(z,\bar z)$ to get:
$$
{\cal I}(\z;k) = 
K(u_1,\z_2,u_3,\z_4;k) 
K(u_1,\z_2,u_3,\z_4;k) C(s,t,u),
$$
where
\bea
\nonumber
&&
K(u_1,\z_2,u_3,\z_4;k) = 
2^{-4} \biggl\{\, \frac{t}{2} {\bar u}_1 \G \z_2 \G(k_3+k_4) 
\G \z_4 u_3 \\
\nonumber
&&
\qquad \qquad \qquad \qquad \qquad \quad
+\frac{s}{2}{\bar u}_1 \G \z_4 \G(k_2+k_3) \G \z_2 u_3 
\biggr\}.
\eea
Now one can recognize in $K(u_1,\z_2,u_3,\z_4;k)$ the standard open string 
kinematical factor of the superstring theory (see \cite{S}).

\subsection{fermion+fermion $\to$ fermion+fermion}
Finally, we consider the kinematical factor corresponding to two fermions in 
the initial and final states.
Our first task is to decompose $T^{\a_1\a_2\a_3\a_4ij}(z)$ into $SO(8)$
invariant rank six spin-tensors. This decomposition is given by
\bea
\nonumber
&&T^{\a_1\a_2\a_3\a_4ij}(z)= \\
\nonumber
&&=
\frac14 {\g^{[ik]}}_{\a_1\a_2} {\g^{[kj]}}_{\a_3\a_4} C_1(z) \\
\nonumber
&&
+\frac12 {\g^{[ij]}}_{\a_3\a_4} \d_{\a_1 \a_2} C_2(z)
+\frac12 {\g^{[ij]}}_{\a_2\a_4} \d_{\a_1 \a_3} C_3(z)
+\frac12 {\g^{[ij]}}_{\a_2\a_3} \d_{\a_1 \a_4} C_4(z) \\
\nonumber
&&
+\frac12 {\g^{[ij]}}_{\a_1\a_4} \d_{\a_2 \a_3} C_5(z)
+\frac12 {\g^{[ij]}}_{\a_1\a_3} \d_{\a_2 \a_4} C_6(z)
+\frac12 {\g^{[ij]}}_{\a_1\a_2} \d_{\a_3 \a_4} C_7(z) \\
\la{decomp}
&&+\,\,\,\,\,\d_{\a_1\a_4} \d^{\a_2\a_3} \d^{ij} C_{8}(z)
  +\,\,\,\,\,\d_{\a_1\a_3} \d^{\a_2\a_4} \d^{ij} C_{9}(z)
  +\,\,\,\,\,\d_{\a_1\a_2} \d^{\a_2\a_4} \d^{ij} C_{10}(z).
\eea
All other $SO(8)$ invariant spin-tensors can be expressed in terms of
linear combinations of spin-tensors from (\ref{decomp}) and therefore are not
linearly independent. To see this, first note that the most general expression 
for such spin-tensor should be at most fourth order in $\g$'s. Indeed,
a term which is of higher than fourth order in $\g$'s and which has only two
vector indices, namely $i$ and $j$, must contain contractions like
$\sum_{k,l} {\g^{[kl]}}_{\a \dot b} {\g^{[kl]}}_{\dot c \dot d}$ where
$\a,\dot b,\dot c,\dot d$ are chosen from $\a_1,\a_2,\a_3,\a_4$. However,
this contraction is just a liner combination of Kronecker deltas as
follows from the identity:
\be
\sum_{k,l} {\g^{[kl]}}_{\a \dot b} {\g^{[kl]}}_{\dot c \dot d} =
8 \d_{\a \dot c} \d_{\dot b \dot d} - 8 \d_{\a \dot d} \d_{\dot b \dot c}.
\la{idty_4}
\ee
However, in (\ref{decomp}) we could have included spin-tensors which are
fourth order in $\g$'s and which are obtained from
${\g^{[ik]}}_{\a_1\a_2} {\g^{[kj]}}_{\a_3\a_4}$ by permuting
spinor indices $\a_1,\a_2,\a_3,\a_4$. Nonetheless, with the account of the
identity
\be
\la{identity5}
{(\g^i \g^k)}_{\a \dot b} {(\g^k \g^j)}_{\dot c \dot d} =
{(\g^k \g^j)}_{\a \dot d} {(\g^i \g^k)}_{\dot b \dot c}
+2 \d_{\a \dot c} {(\g^i \g^j)}_{\dot b \dot d}
+2 \d_{\a \dot b} {(\g^i \g^j)}_{\dot c \dot d}
\ee
it becomes clear that there is only one independent spin-tensor containing
all four $\g$'s and it is represented by the first term in (\ref{decomp}).
This identity is a direct consequence of (\ref{identity2}).
By using the $SU(4) \times U(1)$ basis, we fix all functions $C_i(z)$ and 
their relative phases. The final answer for $T^{\a_1\a_2\a_3\a_4ij}(z)$ is 
given by the following expression
\bea
\nonumber
&&
T^{\a_1\a_2\a_3\a_4ij}(z) =
\frac{n_0^{-\frac12} {(N-n_0)}^{-\frac12}
n_\infty^{-\frac12} {(N-n_\infty)}^{-\frac12}}{n_\infty-n_0}  \\
\nonumber
&&\times \left\{\,\,-
\frac{{(\g^k\g^i)}_{\a_1\a_2}{(\g^k\g^j)}_{\a_3\a_4}}{4}
\frac{(N-n_0)(n_\infty-n_0)(x-\alpha_1)(x-\alpha_2)}{\5\6\7\8\9} \right. \\
\nonumber
&&\,\,\,\,\,\,\,\,-
\frac{{(\g^i\g^j)}_{\a_3 \a_4} \d_{\a_1\a_2}}{2}
\frac{n_0(N-n_0)}{\5\6\7\9} \\
\nonumber
&&\,\,\,\,\,\,\,\,-
\frac{{( \g^i\g^j)}_{\a_2 \a_4} \d_{\a_1\a_3}}{2}
\frac{n_0(N-n_0)}{\5\7\9} \\
\nonumber
&&\,\,\,\,\,\,\,\,-
\frac{{(\g^i\g^j)}_{\a_2 \a_3} \d_{\a_1\a_4}}{2}
\frac{n_0(N-n_\infty)}{\5\7\8\9} \\
\nonumber
&&\,\,\,\,\,\,\,\,-
\frac{{(\g^i\g^j)}_{\a_1 \a_4} \d_{\a_2\a_3}}{2}
\frac{(N-n_0)(n_\infty)}{\5\7\8} \\
\nonumber
&&\,\,\,\,\,\,\,\,-
\frac{{(\g^i\g^j)}_{\a_1 \a_3} \d_{\a_2\a_4}}{2}
\frac{(N-n_0)(N-n_\infty)}{\6\7\8\9} \\
\nonumber
&&\,\,\,\,\,\,\,\,+
\frac{{(\g^i\g^j)}_{\a_1 \a_2} \d_{\a_3\a_4}}{2}
\frac{(N-n_0)(n_\infty-n_0)}{\5\6\7} \\
\nonumber
&&\,\,\,\,\,\,\,\,+
\d^{ij} \d_{\a_1\a_4}\d_{\a_2\a_3}
\frac{n_0(N-n_\infty)\6}{\5\7\8 (x-\alpha_1)(x-\alpha_2)} \\
\nonumber
&&\,\,\,\,\,\,\,\,-
\d^{ij} \d_{\a_1\a_3}\d_{\a_2\a_4}
\frac{n_0(N-n_0)(N-n_\infty)\6}{(n_\infty-n_0)\7\9(x-\alpha_1)(x-\alpha_2)} \\
\nonumber
&&\,\,\,\,\,\,\,\,+ \left.
\d^{ij} \d_{\a_1\a_4}\d_{\a_2\a_3}
\frac{n_0n_\infty(N-n_\infty)}{(n_\infty)\6\7\9(x-\alpha_1)(x-\alpha_2)}
\,\,\right\}.
\eea
The contraction of $T^{\a_1 \a_2 \a_3 \a_4 ij}(z)$ with 
$\l \t_i \t_j \r(z)$ 
is most conveniently performed if we express ${\l \t_i \t_j\r}_k$ in the 
form
\bea
\nonumber
&&
{\l \t_i \t_j \r}_k=
\left(\frac{\5}{n_0}k_1^i-\frac{1}{N-n_0}k_2^i+\frac{\6}{n_\infty}k_3^i\right)
\left(\6 k_1^i - \5 k_3^i+ \frac{n_0-n_\infty}{N-n_\infty}{\9} k_4^i\right) \\
\nonumber
&&
\equiv \t_{ab} k_a^i k_b^j,
\eea
obtained from (\ref{tt}) by using the momentum conservation
law: $\k_1+\k_2+\k_3+\k_4=0$. Since the first
term in $\l \t_i \t_j \r$ contains $\d^{ij}$ and its contraction with
${(\g^k \g^i)}_{\a_1\a_2} {(\g^k\g^j)}_{\a_3\a_4}$ will produce terms which are lower than fourth order in $\g$'s and which at present do not interest us.
So, consider contracting the spin-tensor
${(\g^k \g^i)}_{\a_1\a_2} {(\g^k\g^j)}_{\a_3\a_4}$, i.e. the first term
in (\ref{decomp}), with ${\l \t_i \t_j \r}_k$ and fermionic polarizations
$u_i^{\a_i}$:
\bea
\nonumber
&&
-{(\g^k\g^i)}_{\a_1\a_2} {(\g^k\g^j)}_{\a_3\a_4} u_1^{\a_1} u_2^{\a_2}
u_3^{\a_3} u_4^{\a_4} {\l \t_i \t_j \r}_k = \\
\nonumber
&&=
-\frac14 {\bar u}_1 \G^k \G^i \G^+ u_2
{\bar u}_3 \G^k \G^j \G^+ u_4 {\l \t_i \t_j \r}_k = \\
\nonumber
&&=
-\frac14 {\bar u}_1 \G^{\mu} \G^{\rho} \G^+ u_2
{\bar u}_3 \G_{\mu} \G^{\s} \G^+ u_4 \t_{24} k_2^{\rho} k_4^{\s} 
-
\frac14 {\bar u}_1 \G^{\mu} \G^{\rho} \G^+ u_2
{\bar u}_3 \G_{\mu} \G^{\s} \G^+ u_4 \t_{21} k_2^{\rho} k_1^{\s} \\
\la{T}
&& -
\frac14 {\bar u}_1 \G^{\mu} \G^{\rho} \G^+ u_2
{\bar u}_3 \G_{\mu} \G^{\s} \G^+ u_4 \t_{31} k_3^{\rho} k_1^{\s} 
- 
\frac14 {\bar u}_1 \G^{\mu} \G^{\rho} \G^+ u_2
{\bar u}_3 \G_{\mu} \G^{\s} \G^+ u_4 \t_{34} k_3^{\rho} k_4^{\s}.
\eea
Here again we used the property of ${\l \t_i \t_j \r}_k$, namely (\ref{tau+}),
the nilpotency of $\G^+$ and the fact that $u$ satisfies the Dirac equation.
After commuting $\G^{\rho}$ and $\G^{\s}$ through $\G^+$ and imposing the
Dirac equation, the first term in (\ref{T}) becomes
\bea
\la{ft}
&&
-\frac14 {\bar u}_1 \G^{\mu} \G^{\rho} \G^+ u_2
{\bar u}_3 \G_{\mu} \G^{\s} \G^+ u_4 \t_{24} k_2^{\rho} k_4^{\s}
=
-{\bar u}_1 \G^{\mu} u_2 {\bar u}_3 \G_{\mu} u_4
\t_{24} k_2^+ k_4^+ .
\eea
In order to make use of the Dirac equation in the remaining three terms of
(\ref{T}) we are in need of the identity
\bea
\nonumber
&&
{\bar u}_1 \G^{\mu} \G^{\rho} \G^+ u_2
{\bar u}_3 \G_{\mu} \G^{\s} \G^+ u_4 =
-{\bar u}_1 \G^{\mu} \G^{\s} \G^+ u_4
{\bar u}_2 \G^+ \G^{\rho} \G_{\mu} u_3 \\
\nonumber
&&
-4{\bar u}_1 \G^{\mu} u_3 {\bar u}_4 \G_{\mu} u_2
\eta^{\rho +} \eta^{\s +}
-2{\bar u}_1 \G^{\mu} u_3 {\bar u}_4 (\G^{\s} \G^+ \G_{\mu}
\eta^{\rho +} + \G_{\mu} \G^+ \G^{\rho} \eta^{\s +}) u_2,
\eea
which allows one to place $\G^{\rho}$ next to $u_2$ (or $u_3$) when
it is contracted with $k_2^{\rho}$ (or $k_3^{\rho}$) thereby making it possible
to impose the Dirac equation. This identity just like (\ref{identity5}) is
a direct consequence of (\ref{identity2}). As a result of this procedure and 
with the account of (\ref{T}) and (\ref{ft}) we obtain:
\bea
\nonumber
&&
T^{\a_1 \a_2 \a_3 \a_4}(z) u_1^{\a_1} u_2^{\a_2} u_3^{\a_3} u_4^{\a_4} =
\frac{(N-n_0)(n_\infty-n_0)}{N^2} \frac{(x-\alpha_1)(x-\alpha_2)}{\5\6\7\8} \\
\nonumber
&& \times
\left\{
-{\bar u}_1 \G^{\mu} u_3 {\bar u}_4 \G_{\mu} u_2
 \frac{N-n_0}{n_\infty-n_0} \frac{\6\7}{\9}
+ {\bar u}_1 \G^{\mu} u_2 {\bar u}_3 \G_{\mu} u_4
\right\}.
\eea
Substituting this result into eq.(\ref{I}) we arrive at the
expression for ${\cal I}(\z;k)$
\bea
\nonumber
&&
{\cal I}(\z;k) =
\int d^2z
\left|z\right|^{\frac{1}{2}k_1k_4-2}
\left|1-z\right|^{\frac{1}{2}k_3k_4-2} 
T[u_1,u_2,u_3,u_4](z) 
T[u_1,u_2,u_3,u_4](\bar{z}),
\eea
where
\bea
\nonumber
&&
T[u_1,u_2,u_3,u_4](z) =
\frac{1-z}{4}{\bar u}_1 \G^{\mu} u_3 {\bar u}_4 \G_{\mu} u_2
+
\frac14 {\bar u}_1 \G^{\mu} u_2 {\bar u}_3 \G_{\mu} u_4 \\
\nonumber
&&=
-\frac{1-z}{4} {\bar u}_2 \G^{\mu} u_3 {\bar u}_1 \G_{\mu} u_4
+
\frac{z}{4} {\bar u}_1 \G^{\mu} u_2 {\bar u}_4 \G_{\mu} u_3 .
\eea
Finally, we perform the integration over the sphere $(z,\bar z)$ to get:
$$
{\cal I}(\z;k) = K(u_1,u_2,u_3,u_4;k) K(u_1,u_2,u_3,u_4;k) C(s,t,u),
$$
where
\bea
\nonumber
&&
K(u_1,u_2,u_3,u_4;k) = 2^{-4} \biggl\{ 
-\frac{s}{2} {\bar u}_2 \G^{\mu} u_3 {\bar u}_1 \G_{\mu} u_4
+
\frac{t}{2} {\bar u}_1 \G^{\mu} u_2 {\bar u}_4 \G_{\mu} u_3 \biggr\}.
\eea
We recognize in $K(u_1,u_2,u_3,u_4;k)$ the standard open string kinematic 
factor of the superstring theory (see \cite{S}). For the sake of 
completeness below we provide the kinematical factor corresponding to
four massless vector particles which was calculated in \cite{AF2}.

\subsection{vector particle + vector particle $\to$ vector particle
+ vector particle}
The four graviton scattering amplitude was found in \cite{AF2} and is 
equal to
$$
A(1,2,3,4) = \lambda^2 2^{-8} {\cal I}(\z;k),
$$
where
$$
{\cal I}(\z;k) = K(\z_1,\z_2,\z_3,\z_4;k) K(\z_1,\z_2,\z_3,\z_4;k) C(s,t,u)
$$
and
\bea
\nonumber
&& K(\z_1,\z_2,\z_3,\z_4;k) =  2^{-2} \biggl\{ 
-\frac14 \left( st \z_1 \cdot \z_3 \z_2\cdot \z_4
 + su \z_2\cdot\z_3 \z_1\cdot\z_4 + tu \z_1\cdot\z_2 \z_3\cdot\z_4 \right) \\
\nonumber
&& \qquad 
+\frac{s}{2} \left(
\z_1\cdot k_4\z_3\cdot k_2\z_2\cdot\z_4+
\z_2\cdot k_3\z_4\cdot k_1\z_1\cdot\z_3+
\z_1\cdot k_3\z_4\cdot k_2\z_2\cdot\z_3+
\z_2\cdot k_4\z_3\cdot k_1\z_1\cdot\z_4 \right) \\
\nonumber
&&\qquad + \frac{t}{2} \left(
\z_2\cdot k_1\z_4\cdot k_3\z_3\cdot \z_1+
\z_3\cdot k_4\z_1\cdot k_2\z_2\cdot \z_4+
\z_2\cdot k_4\z_1\cdot k_3\z_3\cdot \z_4+
\z_3\cdot k_1\z_4\cdot k_2\z_1\cdot \z_2
\right) \\
\nonumber
&&\qquad + \frac{u}{2} \left(
\z_1\cdot k_2\z_4\cdot k_3\z_3\cdot \z_2+
\z_3\cdot k_4\z_2\cdot k_1\z_1\cdot \z_4+
\z_1\cdot k_4\z_2\cdot k_3\z_3\cdot \z_4+
\z_3\cdot k_2\z_4\cdot k_1\z_1\cdot \z_2
\right) \biggr\}. 
\eea

\section{Conclusion}
In this paper we obtained kinematical factors and therefore scattering 
amplitudes for all massless particles of type IIA superstrings directly from 
the interacting $S^N \R^8$ orbifold sigma model. Our kinematical factors
showed to coincide with those obtained in the framework of the superstring
theory. This provides further evidence of
the duality between the Yang-Mills theory in the IR limit and the superstring
theory in the week coupling limit.

In computing the scattering amplitudes we did not impose any kinematic
restrictions on momenta and polarizations of particles. Nevertheless, the
obtained kinematical factors which define scattering amplitudes exhibit
manifest Lorentz invariance even at finite $N$.
All dependence on $N$ was absorbed into the light-cone momenta $k^+$.

Moreover, if one restores the dependence on the radius $R_-$ of the
compactified direction $x_-$ (remind that $N$ was identified with $R_-$)
then any dependence on $N$ disappears.  Since the $S^N\R^8$ orbifold model
can be embedded into the  $S^\infty\R^8$ orbifold model, this suggests that
the latter might have a deformed (quantum) Lorentz symmetry realized
in the space of the twist fields $\S_{(n)}^{\dot \mu}$. The
deformation parameter seems to be identified with $exp(2\pi i/R_-)$.

\noindent{\bf ACKNOWLEDGMENTS}

The authors thank L.O.Chekhov and A.A.Slavnov for valuable discussions. 
The work of G.A. was supported by the Cariplo Foundation for
Scientific Research and in part by the RFBI grant N96-01-00608,
and the work of S.F. was supported by the U.S. Department of Energy under
grant No. DE-FG02-96ER40967 and in part by the RFBI grant N96-01-00551.
\setcounter{section}{0}
\appendix{}
\setcounter{equation}{0}
We use the following representation of $\g$-matrices
satisfying the relation
$$
\g^i (\g^j)^T+\g^j (\g^i)^T=2\d^{ij}I
$$
\bea
\nonumber
\begin{array}{ll}
\gamma^1=
\left(\begin{array}{rr}1&0\\0&1\end{array}\right)\otimes
\left(\begin{array}{rr}0&1\\1&0\end{array}\right)\otimes
\left(\begin{array}{rr}0&1\\-1&0\end{array}\right)  &
\gamma^2=
\left(\begin{array}{rr}0&1\\-1&0\end{array}\right)\otimes
\left(\begin{array}{rr}0&1\\-1&0\end{array}\right)\otimes
\left(\begin{array}{rr}0&1\\-1&0\end{array}\right)  \\
\gamma^3=1 &
\gamma^4=
\left(\begin{array}{rr}0&1\\-1&0\end{array}\right)\otimes
\left(\begin{array}{rr}1&0\\0&1\end{array}\right)\otimes
\left(\begin{array}{rr}1&0\\0&-1\end{array}\right) \\
\gamma^5=
\left(\begin{array}{rr}1&0\\0&1\end{array}\right)\otimes
\left(\begin{array}{rr}1&0\\0&-1\end{array}\right)\otimes
\left(\begin{array}{rr}0&1\\-1&0\end{array}\right)
&
\gamma^6=-
\left(\begin{array}{rr}0&1\\-1&0\end{array}\right)\otimes
\left(\begin{array}{rr}1&0\\0&1\end{array}\right)\otimes
\left(\begin{array}{rr}0&1\\1&0\end{array}\right) \\
\gamma^7=
\left(\begin{array}{rr}1&0\\0&-1\end{array}\right)\otimes
\left(\begin{array}{rr}0&1\\-1&0\end{array}\right)\otimes
\left(\begin{array}{rr}1&0\\0&1\end{array}\right)
&
\gamma^8=
\left(\begin{array}{rr}0&1\\1&0\end{array}\right)\otimes
\left(\begin{array}{rr}0&1\\-1&0\end{array}\right)\otimes
\left(\begin{array}{rr}1&0\\0&1\end{array}\right).
\end{array}
\eea
\bea
\la{G}
&&\G^0 =  \s_2 \otimes 1_{16}, \\
\nonumber
&&\G^i = i\s_3 \otimes \left(\begin{array}{cc}0&{\g}^i \\{(\g^i)}^T&0
\end{array}\right), \quad i = 1, \ldots, 8, \\
\nonumber
&&\G^9 = i\s_1 \otimes 1_{16}, \\
\nonumber
\nonumber
&& \G_{11} = \G^0\G^1\ldots\G^9 = \s_3 \otimes \s_3 \otimes 1_8 .
\eea
By definition,
\bea
\la{idty1}
&&\U^{[\mu \nu \lambda \rho]} = \frac{1}{4!}\sum_{P}(-1)^{P(\mu\nu\lambda\rho)}
\U^{\mu}\U^{\nu}\U^{\lambda}\U^{\rho} = \\
\nonumber
&&
\frac16
(\U^{[\mu\nu]}\U^{[\lambda\rho]}
+\U^{[\lambda\rho]}\U^{[\mu\nu]}
-\U^{[\mu\lambda]}\U^{[\nu\rho]}
-\U^{[\nu\rho]}\U^{[\mu\lambda]}
+\U^{[\mu\rho]}\U^{[\nu\lambda]}
+\U^{[\nu\lambda]}\U^{[\mu\rho]}).
\eea
Here $\U$ can be either $\g$ or $\G$.

In terms of ordinary products of $\U$-matrices
$\U^{[\mu\nu\lambda\rho]}$ is expressed as follows
\bea
\la{identity1'}
&&
\U^{[\mu\nu\lambda\rho]} =
 \U^{\mu}\U^{\nu}\U^{\lambda}\U^{\rho}
-\U^{\lambda}\U^{\rho}\eta^{\mu\nu} + \U^{\nu}\U^{\rho}\eta^{\mu\lambda}
-\U^{\nu}\U^{\lambda}\eta^{\mu\rho} - \U^{\mu}\U^{\rho}\eta^{\nu\lambda}
+\U^{\mu}\U^{\lambda}\eta^{\nu\rho} - \U^{\mu}\U^{\nu}\eta^{\lambda\rho}\\
\nonumber
&&+\eta^{\mu\nu}\eta^{\lambda\rho} - \eta^{\mu \lambda}\eta^{\nu \rho} +
\eta^{\mu \rho}\eta^{\nu \lambda}.
\eea
In $D=10$ $\G$'s satisfy the following equality (see e.g. \cite{GSW}):
\be
\la{identity2}
{(\G^{0} \G^{\mu})}_{mn}{(\G^{0}\G_{\mu})}_{pq} +
{(\G^{0} \G^{\mu})}_{mp}{(\G^{0}\G_{\mu})}_{qn} +
{(\G^{0} \G^{\mu})}_{mq}{(\G^{0}\G_{\mu})}_{np} = 0.
\ee
Here it is assumed that spinor indices have definite chirality.

\appendix{}
\setcounter{equation}{0}
With respect to the $SU(4)\times U(1)$ subgroup
representations ${\bf 8_v}$, ${\bf 8_s}$ and ${\bf 8_c}$
are decomposed as
\bea
\nonumber
{\bf 8_s}\to {\bf 4_{1/2}}+{\bf \bar{4}_{-1/2}}, \qquad
{\bf 8_c}\to {\bf 4_{-1/2}}+{\bf \bar{4}_{1/2}}, \qquad
{\bf 8_v}\to {\bf 6_0}+{\bf 1_1}+{\bf 1_{-1}}.
\eea
The corresponding basis for the fermions $\theta^a$ and their
spin fields\footnote{see, e.g., \cite{FMS} for a detailed discussion of spin
fields} $\S^{\dot{a}}$ and $\S^{i}$
consistent with this decomposition is given by
\bea
\nonumber
&&\T^A=\frac{1}{\sqrt{2}}(\theta^A+i\theta^{A+4}), \qquad
\T^{\bar A}=\frac{1}{\sqrt{2}}(\theta^A-i\theta^{A+4}), \\
\nonumber
&&\CS^{\dot{A}}=\frac{1}{\sqrt{2}}(\S^{\dot{A}}+i\S^{\dot{A}+4}), \qquad
\CS^{\dot{\bar{A}}}=\frac{1}{\sqrt{2}}(\S^{\dot{A}}-i\S^{\dot{A}+4}), \\
\nonumber
&&S^{A}=\frac{1}{\sqrt{2}}(\S^{2A-1}+i\S^{2A}), \qquad
S^{\bar{A}}=\frac{1}{\sqrt{2}}(\S^{2A-1}-i\S^{2A}),
\eea
where $A=1,\ldots ,4$. Note that the spin fields $\S^4$ and
$\S^{\bar{4}}$ transform as ${\bf 1_1}$ and ${\bf 1_{-1}}$ respectively.

Bosonization of the fermions and their twist fields
up to cocycles is realized in terms of four bosonic fields $\phi^A$ as
\bea
\nonumber
\T^A=\e^{iq_B^A\phi^B}, \qquad
\CS^{\dot{A}}=\e^{iq^{\A}_{B}\phi^B},\qquad
S^{A}=\e^{i\phi^A},
\eea
where the weights of the spinor representations ${\bf 8_s}$ and ${\bf 8_c}$
are given by
\bea
\nonumber
&&\q^1=\frac{1}{2}(-1,-1,1,1);~~~
\q^2=\frac{1}{2}(-1,1,-1,1);~~~
\q^3=\frac{1}{2}(1,-1,-1,1);~~~
\q^4=\frac{1}{2}(1,1,1,1); \\
\la{weight}
&&
\q^{\dot 1}=\frac{1}{2}(-1,1,1,1);~~~
\q^{\dot 2}=\frac{1}{2}(-1,-1,-1,1);~~~
\q^{\dot 3}=\frac{1}{2}(1,1,-1,1);~~~
\q^{\dot 4}=\frac{1}{2}(1,-1,1,1).
\eea
The Cartan generators of
$SU(4)\times U(1)$ in the bosonized form look as $H^A=i\partial \phi^A$.

Bosonization of the fermions of the orbifold model
is achieved by introducing $4N$ bosonic fields and reads as
$$
\T_I^A(z)=\e^{iq_B^A\phi^B_I(z)}.
$$
Twist fields $\s_g$ creating twisted sectors for the fields
$\phi^A_I(z)$ are introduced in the same manner as in Sec.2.2.
The spin twist fields of the orbifold model can be realized as
\bea
\la{bosor}
&&\CS_{(n)}^{\dot{A}}(z)=
\e^{\frac{i}{n}\sum_{I\in (n)}q^{\A}_{B}\phi^B_I(z)}\s_{(n)}(z)=
\s_{(n)}[\q^{\A}](z),
\\
\nonumber
&&S_{(n)}^{A}(z)=\e^{\frac{i}{n}\sum_{I\in (n)}\phi^A_I(z)}\s_{(n)}(z)=
\s_{(n)}[\eb^{A}](z),
\eea
where $\eb^{A}$ is a weight vector of ${\bf 8_v}$ with
components $\d^{A}_{B}$.


\end{document}